\documentclass[]{emulateapj}
\usepackage{apjfonts}
\pdfoutput=1
\newcommand{\beq}{\begin{equation}}
\newcommand{\eeq}{\end{equation}}
\newcommand{\bea}{\begin{eqnarray}}
\newcommand{\eea}{\end{eqnarray}}

\begin{document}
\title{Supernovae Powered by Collapsar Accretion in Gamma-Ray Burst Sources}

\author{Milo\v s Milosavljevi\'c \altaffilmark{1}, 
Christopher C. Lindner \altaffilmark{2},
Rongfeng Shen \altaffilmark{3},
and
Pawan Kumar}
\affil{Department of Astronomy, University of Texas, 1 University Station C1400, Austin, TX 78712.}
\altaffiltext{1}{Texas Cosmology Center, University of Texas, 1 University Station C1400, Austin, TX 78712.}
\altaffiltext{2}{NSF Graduate Research Fellow.}
\altaffiltext{3}{Currently at Department of Astronomy, University of Toronto, 50 St. George St., Toronto, Ontario M5S 3H4, Canada.}

\righthead{SUPERNOVAE POWERED BY COLLAPSAR ACCRETION}
\lefthead{}

\begin{abstract}

The association of long-duration gamma-ray bursts (LGRBs) with Type Ic supernovae presents a challenge to supernova explosion models.  In the collapsar model for LGRBs, gamma rays are produced in an ultrarelativistic jet launching from the magnetosphere of the black hole that forms in the aftermath of the collapse of a rotating progenitor star.  The jet is collimated along the star's rotation axis, but the concomitant luminous supernova should be relatively---though certainly not entirely---spherical, and should synthesize a substantial mass of $^{56}$Ni.  Our goal is to provide a qualitative assessment of the possibility that accretion of the progenitor envelope onto the black hole, which powers the LGRB, could also deposit sufficient energy and nickel mass in the envelope to produce a luminous supernova. For this, the energy dissipated near the black hole during accretion must be transported outward, where it can drive a supernova-like shockwave.  Here we suggest that the energy is transported by convection and develop an analytical toy model, relying on global mass and energy conservation, for the dynamics of stellar collapse.  The model suggests that a $\sim 10,000\,\textrm{km}\,\textrm{s}^{-1}$ shock can be driven into the envelope and that $\sim 10^{51}\,\textrm{erg}$ explosions are possible.  The efficiency with which the accretion energy is being transferred to the envelope is governed by the competition of advection and convection at distances $\sim 100-1,000\,\textrm{km}$ from the black hole and is sensitive to the values of the convective mixing length, the magnitude of the effective viscous stress, and the specific angular momentum of the infalling envelope.  Substantial masses of $^{56}$Ni may be synthesized in the convective accretion flow over the course of tens of seconds from the initial circularization of the infalling envelope around the black hole.  The synthesized nickel is convectively mixed with a much larger mass of unburned ejecta.

\keywords{ accretion, accretion disks --- black hole physics --- gamma rays: bursts --- nuclear reactions, nucleosynthesis, abundances --- supernovae: general }

\end{abstract}

\section{Introduction}
\label{sec:intro}
\setcounter{footnote}{0}

A growing number of long-duration gamma-ray bursts (LGRBs) are being discovered in association with Type Ic supernovae \citep{Galama:98,Galama:00,Reichart:99,Bloom:02,DellaValle:03,DellaValle:06,Garnavich:03,Hjorth:03,Kawabata:03,Stanek:03,Matheson:03,Malesani:04,Campana:06,Mirabal:06,Modjaz:06,Pian:06,Chornock:10,Cobb:10,Starling:11}, yet on the basis of the non-detection of late-time radio emission in a sample of Type Ic supernovae, \citet{Podsiadlowski:04}  and \citet{Soderberg:06b} inferred that less than $10\%$ of all Type Ic supernovae are associated with standard LGRBs.  The process producing LGRBs and their concomitant supernovae remains a subject of debate \citep[][and references therein]{Woosley:06b}.  In the collapsar model for LGRBs \citep{Woosley:93}, the gamma rays are produced in an ultrarelativistic jet launching from the magnetosphere of the black hole that forms in the aftermath of the collapse of a rotating progenitor. The jet is powered by a continuous infall and disklike accretion of the progenitor star's interior.  While the collapsar model seems to successfully explain the power and duration of LGRBs, it is not clear at present whether it naturally gives rise to a supernova-like stellar explosion.  It has been argued that a ``wind'' outflowing from the nonradiative parts of the collapsar disk may convey sufficient energy to the stellar envelope for an explosion, and that $^{56}$Ni is synthesized in the wind to later produce an optically-bright supernova \citep[e.g.,][]{MacFadyen:99,MacFadyen:03,Pruet:03,Pruet:04,Kohri:05}.  The mechanics of energy transfer from the disk to the supernova ejecta and its implications for nickel synthesis remain open problems and are the subject of the present study.

Here we provide a crude assessment of the possibility that the accretion onto the black hole that powers the LGRB might also deposit sufficient energy in the progenitor envelope to produce a supernova. For this, the accretion energy dissipated near the black hole must be transported to exterior mass coordinates of the star.  We hypothesize that the energy is transported by convection to energize the outward moving shock, as was originally suggested by \citet{Narayan:01}. Our two-dimensional simulations of collapsar accretion \citep{Lindner:10}, in which we simulated only relatively large radii $(r>500\,\textrm{km})$ and did not incorporate neutrino and nuclear physics, corroborate the crucial role of convection.  Our approach in the present work differs from existing assessments of the viability of collapsar supernovae \citep[e.g.,][]{Kohri:05} in that we attempt to sketch out the global structure of the flow by incorporating disk accretion, convective energy transport, shock dynamics, and long-term stellar infall (the latter having been studied in \citealt{Kumar:08a,Kumar:08b}) in a single toy model.  Particularly important are the non-Keplerian nature of the accretion flow and the precise form of the viscous torque in a pressure-supported regime; the latter aspects seem to have been neglected in existing treatments of collapsar disks but came to light in the 2.5D simulations of \citet{Lindner:10}.

The toy model allows us to investigate how the shock expansion interferes with the rate with which the infalling material is accreting onto the black hole.  This differs from the approaches that implicitly postulate, as in the advection-dominated accretion flow (ADAF) paradigm \citep[e.g.,][]{Narayan:94,Narayan:95,Blandford:99}, that the inflow occurs in the equatorial region and is spatially separated from a non-interfering wind that carries mass and energy into an axial outflow cone.   Our treatment is similar in spirit to the toy model constructed by \citet{Janka:01}, following in the footsteps of \citet{Bethe:90,Bethe:93a,Bethe:93b,Bethe:95,Bethe:96,Bethe:97}, to assess conditions for shock revival by neutrino heating in core-collapse supernovae; the key differences include the central roles of rotation and convection in collapsars, and comparatively lower mass accretion rates, longer time scales, and lower fluid densities in the latter systems.

This work is organized as follows.  In Section \ref{sec:shock_wave} we discuss the formation of an accretion shock and the post-shock conditions immediately following shock formation. In Section \ref{sec:adaf_cdaf} we analyze the structure of the inner accretion flow and estimate the luminosity that convection can transport toward the shock wave.  In Section \ref{sec:conservation} we impose global mass and energy conservation and in Section \ref{sec:results} we utilize these to so estimate the shock expansion velocity and the total energy deposited in the stellar envelope.  In Section \ref{sec:nucleosynthesis} we discuss the prospects for $^{56}$Ni production in our model. In Section \ref{sec:conclusions} we present our conclusions and briefly discuss some implications.

\section{The Accretion Shock}
\label{sec:shock_wave}

The formation of the black hole in stellar core collapse might be preceded by a failed explosion resulting from a bounce and subsequent shock revival by neutrino heating \citep[][see also \citealt{MacFadyen:01}]{Bethe:85}, or by an ejection of a magnetic field from a magnetized proto-neutron star \citep[e.g.,][]{BisnovatyiKogan:71,Wheeler:00,Thompson:04,Bucciantini:07,Burrows:07,Dessart:08}.  The magnetic outflow may be too axially collimated to produce a standard supernova explosion \citep{Bucciantini:08,Bucciantini:09}.  Here we ignore the possibility of any type of explosion preceding the collapse into a black hole, and assume that during the first few seconds from the collapse of the stellar core, an unshocked stellar envelope accretes quasi-radially onto the black hole. 

For a few solar mass black hole, direct unshocked accretion through the event horizon of the black hole is possible during an initial interval measuring in the tens of seconds, while the specific angular momentum of the infalling shells is in the range $\ell \lesssim 10^{16}\, (M_{\rm BH}/M_\odot) \,\textrm{cm}^2\,\textrm{s}^{-1}$ ($M_{\rm BH}$ is the black hole mass and the critical specific angular momentum decreases with increasing black hole spin for prograde accretion).   Because the specific angular momentum of the innermost stellar shells increases outward, unshocked accretion takes place as shells at mass coordinates $\sim (3-6)M_\odot$ arrive near the black hole in the unmixed and fully mixed pre-supernova models \citep[see, e.g.,][]{Heger:00,Heger:05,MacFadyen:01,Petrovic:05,Woosley:06a}. When the specific angular momentum arriving near the innermost stable orbit around the black hole becomes comparable to the angular momentum of the innermost stable circular orbit (ISCO), the accretion occurs through a ``dwarf'' or ``mini''-disk \citep{Lee:06,Zalamea:09}; during this transitional period, the black hole could acquire rapid rotation and accretion takes place at the rate $\dot M\sim (0.1-0.2) \,M_\odot\,\textrm{s}^{-1}$.  At some time $t_{\rm sh}$ that depends on the stellar mass and rotational profile, the flow crossing the ISCO becomes subsonic and a quasispherical accretion shock wave forms around the black hole \citep[e.g.,][]{MacFadyen:99,Lee:06,Nagataki:07,LopezCamara:09,Lindner:10}.  

The initial radius of the shock, $r_{\rm sh}$, is just larger than $r_{\rm ISCO}\sim 5-50\,\textrm{km}$, where the latter depends on the mass and angular momentum of the black hole.  If the shock proceeds to travel outward, the rate with which material accretes onto the black hole drops rapidly \citep{Lindner:10}, e.g., by a factor of $10$ or more during the first second from the inception of the shock.  Dynamics of the shock wave is governed by the rate of stellar envelope infall and the conditions in the rotating downstream fluid.  Besides the heating at the accretion shock, the fluid is heated by the dissipation of magnetohydrodynamic turbulence driven by the magnetorotational instability \citep[MRI, e.g.,][and references therein]{Thompson:05}.  The fluid cools by neutrino emission, primarily through pair annihilation and pair capture onto nucleons (the Urca process).  The fluid also cools through the disintegration of nuclei into helium and free nucleons as it passes the shock and accretes toward the black hole.  The latter process can be reversible, as the energy consumed in disintegration can be recovered if the free nucleons and helium end up getting transported by convection (``dredged up'') to radii with lower entropies where they can recombine into heavier nuclei.  

The character of the flow is sensitive to the relative magnitude of the cooling and heating rates.  When the cooling is comparable to the heating, the flow collapses into a rotationally-supported, neutrino-cooled accretion disk \citep[e.g.,][]{MacFadyen:99,Popham:99,Narayan:01,DiMatteo:02,Kohri:02,Janiuk:04,Setiawan:04,Kohri:05,Lee:05,Chen:07,Kawanaka:07}. The flow accretes onto the black hole through a thin disk when the accretion rate is higher than a minimum value that depends on the viscous stress-to-pressure ratio $\alpha$ and the black hole's spin parameter $a$; this is because disks with a larger $\alpha$ are less dense and cooler.  In disks that are optically thin to neutrinos, for a given $\alpha$, neutrino cooling dominates disk thermodynamics at relatively high accretion rates and small radii; at lower accretion rates or larger radii, the flow becomes geometrically thick and non-radiative.  For example, \citet{Chen:07} found that for $\alpha=0.01$, a neutrino cooled disk can be present around a $M_{\rm BH}=3\,M_\odot$ black hole for $\dot M\gtrsim (10^{-4}-10^{-3})\,M_\odot\,\textrm{s}^{-1}$, but for $\alpha=0.1$, the accretion rate must be $\dot M\gtrsim (0.02-0.1)\,M_\odot\,\textrm{s}^{-1}$ for a thin disk to be present.\footnote{If the volumetric neutrino cooling rate is $Q_\nu\propto \rho^A\, T^B$, where $(A,B)=(0,9)$ for pair annihilation and $(A,B)=(1,6)$ for Urca, and if, for the purpose of illustration, radiation and low-density relativistic pairs dominate the pressure at the thin-to-thick transition radius $r_\nu$ and the gravitational field is Newtonian, then it is straightforward to show that for fixed $\dot M$, we have $r_\nu\propto \alpha^{2(4A+B)/(32-12A+5B)}$, which gives $r_\nu\propto \alpha^{-18/13}$ for pair annihilation and $r_\nu\propto \alpha^{-2}$ for Urca.  The value of $\alpha$ must be smaller than a critical maximum value if $r_\nu$ is to be larger than $r_{\rm ISCO}$, as required for neutrino-cooled, think-disk accretion.}  

The true value of $\alpha$ in the regime in which pressure and centrifugal forces are of the same order, and where the fluid, as we shall see, is convective, is not known.  If the flow is convective, then the convection may contribute to the buildup of the magnetic stress \citep{Balbus:02,Igumenshchev:02,Igumenshchev:03,Christodoulou:03}, and this may motivate a large value of $\alpha$ and a large critical accretion rate required for the presence of a thin, neutrino cooled disk.  If $\alpha\sim 0.1$, which is the value we take as the fiducial for what follows, then the very initial accretion rate drop following shock expansion already brings the accretion rate below the critical rate for efficient neutrino cooling, and the flow is nonradiative and geometrically thick at all radii.\footnote{In our companion 1.5D numerical simulations \citep{Lindner:11}, we estimated the vertical-pressure-scale-height-to-radius ratio and found that in our fiducial model, within $r \lesssim (100-200) \,\textrm{km}$ the ratio dipped below $0.5$, the value characteristic of a geometrically thick flow, but only moderately, to $0.3$. The flow remained relatively geometrically thick at all times.  We attributed the observed moderate thinning of the accretoin flow to the cooling of the flow by the dissociation of helium nuclei into free nucleons.}  The toy model that we present below  will be restricted to this non-neutrino-cooled regime.

It may be worth noting that the picture in which a supersonically infalling flow passes an accretion shock, becomes predominantly rotationally supported, and proceeds to accrete onto a central compact object, either through a neutrino-cooled thin disk, or through a nonradiative thick disk, resonates with the work examining the post-supernova fallback onto a neutron star, or examining the Bondi-Hoyle accretion onto a neutron star embedded in a common envelope  \citep[see, e.g.,][and references therein]{Chevalier:96,Brown:00}.  Because the characteristic accretion rates in these contexts, which are $\sim 1\,M_\odot\,\textrm{yr}^{-1}$, are orders of magnitude below those anticipated in collapsars, the maximum values of $\alpha$ for which neutrino-cooled disk solutions exist are much smaller than those in collapsars. Furthermore, photon diffusion may be relevant in the fallback and common-envelope contexts \citep[see, e.g.,][]{Blondin:86}, but in collapsars, complete photon trapping is a safe assumption.

We work under the assumption that the accretion shock remains quasi-spherical as it traverses the star, and thus, that the thermal ``wind'' produced in the inner accretion disk remains trapped within the surface of the shock so that the wind's energy is distributed quasi-spherically behind the shock, as seen in idealized 2.5D simulations \citep{Lindner:10}.  We can allow for the possibility that a collimated electromagnetic outflow distinct from the thermal wind, such as a jet enveloped by a cocoon of shocked stellar fluid \citep[see, e.g.,][]{Zhang:03,Zhang:04,ZhangW:06,Morsony:07,Wang:08}, is present along the axis of rotation; our analysis should be construed as applying to the equatorial region not occupied by the jet.  Regardless of the presence of the jet, on time scales much shorter than the free fall time from the surface of the star, the axial ``funnel'' region is not empty, and remains overpressured either by the freely falling axial low-angular-momentum material, or by the jet's hot cocoon.  Thermal outflow from the predominantly rotationally-supported central accretion flow launches at oblique angles from the surface of the disk, following ``gyrentropes'' \citep[the surfaces of approximately constant angular momentum, Bernoulli function, and entropy; see][]{Blandford:04}, as is evident in numerous simulations of radiatively-inefficient accretion flows in regions in which the magnetic field is not dynamically important \citep{Stone:99,Igumenshchev:00b,Abramowicz:02,Hawley:02,Proga:03a,Proga:03b,Lindner:10}.  The thermal wind then mixes with the post-shock fluid; in this sense, we think of the oblique thermal wind as a form of convection (or, more adequately, stochastic circulation) with an effective mixing length that can be large and need not be limited by the local pressure scale height.  

At radii $r\gg r_{\rm ISCO}$, the turbulent dissipation rate (due to MRI) is a steeply declining function of radius and this gives rise to strong entropy inversion and convective instability.  The degree of rotational support in the post-shock fluid increases inward \citep{Lindner:10}.  The inner, rotationally-supported torus may, according to the Solberg-H\o iland criterion, be convectively stable in the equatorial direction; instability is still present in a direction inclined relative to the equator, and a fluid element thus transported obliquely, along a gyrentrope, eventually mixes with the denser equatorial fluid.  With this in mind, we develop an effective, spherically averaged picture in which we postulate that convective heat transport proceeds according to the prescription of mixing length theory (MLT) for a non-rotating atmosphere in which the mixing length is interpreted as a parameter that hides the complexity arising from the rotation and vertical stratification. 

The specific angular momentum of the post-shock fluid immediately following shock formation is only slightly larger than that of a circular orbit at $r_{\rm ISCO}$.  The mass of the shocked fluid comprises only a small fraction of the mass of the progenitor star.  Barring an extremely steep pre-collapse radial gradient in the specific angular momentum of the progenitor star, the specific angular momentum in the shocked fluid varies only over a narrow range of values, and is further homogenized by convective mixing.  At the smallest radii, viscous redistribution produces a small, positive gradient in $\ell$ turning over to a small negative gradient at radii at which the viscous time (associated with the MRI stress) becomes comparable to the age of the flow.  At radii where the viscous time is longer than the age of the flow so that no significant viscous angular momentum redistribution could have taken place, the specific angular momentum is a passive scalar transported by convective eddies. We refer the reader to Figure 4 in \citet{Lindner:10}, where the near-radial-independence of the specific angular momentum of the shocked fluid can clearly be seen.

\section{Advection and Convection}
\label{sec:adaf_cdaf}

A fraction of the energy dissipated by the accreting shocked stellar envelope is advected into the black hole; the rest is transported outward by convection and can power an explosion.  Here we study the competition between advection and convection in the innermost segments of the accretion flow and attempt to assess the asymptotic luminosity carried by the shocked envelope.  We work in the spherically averaged picture in which fluid variables depend on the spherical radius $r$ and represent spherical averages over the angular coordinates $\theta$ and $\phi$.  In Section \ref{sec:energetics} we write relations for the transport of internal and total energy in the inner accretion flow, and in Section \ref{sec:support} we discuss the nature of radial force balance in the flow.  In Section \ref{sec:radial_structure_model} we present our toy model for the radial structure of the flow.  In Section \ref{sec:eos} we justify our adoption of a simple equation of state that will serve as basis for our toy models.  In Section \ref{sec:time_scales} we review the several key time scales characterizing the innermost accretion flow. In Section \ref{sec:convection} we argue that convection is an effective energy transport mechanism only down to some minimum radius; at still smaller radii, the energy dissipated by accretion is advected into the black hole.  In Section \ref{sec:luminosity} we provide estimates of luminosity carried by convection. In Section \ref{sec:nuclear} we discuss the impact of nuclear processes, and in Section \ref{sec:summary_advection_convection}, we summarize our conclusions to help us prepare to undertake an analysis of the structure and dynamics of the envelope in Sections \ref{sec:conservation} and \ref{sec:results}.

\subsection{Energetics and Transport}
\label{sec:energetics}

At radii much smaller than the radius of the shock, $r\ll r_{\rm sh}$, where the accretion flow is in a quasi-steady state characterized by a radial force balance and the bulk inward motion is entirely due to viscous accretion, the conservation of internal energy can be written in the form
\beq
\label{eq:energy_conservation}
v_r \rho T\frac{ds}{dr} + \frac{1}{r^2} \frac{d}{dr} [r^2 (F_{\rm conv}+v_r\rho\epsilon_{\rm nuc}+F_{\rm mix} )] = Q_{\rm visc} - Q_\nu ,
\eeq
where $v_r$ is the mass-weighted average radial velocity, $\rho$ is the fluid density, $s$ is the specific entropy, $\epsilon_{\rm nuc}$ is the specific (negative) nuclear binding energy, $F_{\rm conv}$ is the heat flux carried by convection, $F_{\rm mix}$ is the flux of nuclear binding energy due to convective mixing, $Q_{\rm visc}$ is the rate of viscous dissipation that is proportional to the square of the local shearing rate, and $Q_\nu$ is the rate of cooling through neutrino emission under optically thin conditions.   For the purpose of analytic transparency and clarity, we have opted not to carry out a formally self-consistent spherical averaging procedure in which all specific-angular-momentum-dependent terms in equation (\ref{eq:energy_conservation}) and, depending on the symmetries assumed, the forthcoming equations could carry additional numerical factors resulting from the $\theta$-dependence of $\ell(r,\theta,\phi)$.\footnote{For example, if the density $\rho$ is assumed to be spherically symmetric and spherical shells are assumed to rotate rigidly, $\ell\propto\sin^2\theta$, then the ``$\ell$'' appearing in the equations in this section can be interpreted as representing $2/3$ of the maximum, equatorial specific angular momentum, while the terms quadratic in $\ell$ would require an overall correction factor of $6/5$.}

In the part of the flow where the fluid is in radial force equilibrium, the radial motion associated with the viscous angular momentum transport occurs with velocity
\beq
\label{eq:v_r_viscous}
v_r\sim \left(r^2\rho \frac{d\ell}{dr}\right)^{-1}  \frac{d}{dr}\left(r^4\nu\rho \frac{d\Omega}{dr}\right) ,
\eeq
where $\nu$ is the kinematic shear viscosity and $\Omega=\ell/r^2$ is the angular velocity.
This result applies at the radii that are in viscous quasi-equilibrium, i.e., where radial angular momentum transport rate is approximately independent of radius; note also that $v_r(r)$ must be continuous and differentiable at any local extrema of $\ell(r)$. 

The conservation of total energy can be expressed as
\bea
\label{eq:total_energy_conservation}
\frac{1}{r^2} \frac{d}{dr} \left\{r^2 v_r \rho\left[\frac{1}{2} \left(v_r^2+\frac{\ell^2}{r^2}\right) + \frac{\gamma P}{(\gamma-1)\rho} + \epsilon_{\rm nuc} + \Phi \right]
\nonumber\right. \\ \left.
- r^2\rho\nu \ell \frac{d\Omega}{dr}+r^2 (F_{\rm conv}+F_{\rm mix})\right\} =- Q_\nu ,
\eea
where $\gamma$ is the adiabatic index of the shocked fluid and $\Phi$ is the gravitational potential. We work in the approximation in which the gravitational potential and fluid mechanics are nonrelativistic; this is clearly not true close to the black hole, but for our purposes it will suffice that it be a good approximation outside the innermost advective region.  In line with our hypothesis that the post-shock-formation accretion rate drops below the critical value for efficient neutrino cooling, we will assume $Q_\nu\approx 0$.  

In MLT, if global compositional gradients and nuclear composition changes inside convective cells can be ignored, the heat flux carried by convection is 
\beq
\label{eq:convective_flux}
F_{\rm conv}=\frac{1}{4}c_P\rho \left[-\frac{\nabla\Phi}{\rho}\left(\frac{\partial\rho}{\partial T}\right)_P\right]^{1/2} \lambda_{\rm conv}^2 \left(-\frac{T}{c_P}\frac{ds}{dr}\right)^{3/2} ,
\eeq 
where $c_P$ is the specific heat at constant pressure and $\lambda_{\rm conv}$ is the convective mixing length.
The nuclear binding energy is a sum over nuclear species, $\epsilon_{\rm nuc}=\sum_i E_i X_i / m_i$, where $E_i$, $X_i$, and $m_i$ denote, respectively, the binding energies, mass fractions, and nuclear masses of the species.  
We model the mixing of nuclear species in the diffusion approximation \citep[e.g.,][]{Cloutman:76,Kuhfuss:86}
\beq
\label{eq:mixing}
\left[\frac{\partial (\rho X_i)}{\partial t}\right]_{\rm mix}  =  \frac{1}{r^2} \frac{\partial}{\partial r} \left(r^2 \,\frac{1}{3}\chi_{\rm mix} v_{\rm conv} \lambda_{\rm conv} \rho \frac{\partial X_i}{\partial r}\right) ,
\eeq
where 
\beq
\label{eq:convective_velocity} 
v_{\rm conv}\sim \frac{1}{2}\lambda_{\rm conv} \left[\frac{\nabla \Phi}{\rho}\left(\frac{\partial\rho}{\partial T}\right)_P  \frac{T}{c_P} \frac{ds}{dr}\right]^{1/2} ,
\eeq 
is the velocity of convective cells, and $\chi_{\rm mix}\sim 1$ is a dimensionless parameter characterizing the efficiency of convective mixing.\footnote{For a criticism of the application of MLT to mixing, see e.g., \citet{Ventura:98} and references therein.} 
The flux of energy due to the convective mixing can then be obtained by multiplying equation (\ref{eq:mixing}) with $E_i/m_i$ and summing over nuclear species to find
\beq
F_{\rm mix}=-\frac{1}{3}\chi_{\rm mix}v_{\rm conv}\lambda_{\rm conv} \rho \frac{d\epsilon_{\rm nuc}}{dr} .
\eeq 
We assume that the convective motions are subsonic.

\subsection{Radial Force Balance and Viscosity}
\label{sec:support}

If the shocked accretion flow has nearly uniform specific angular momentum, the fractional contribution of rotation to radial force balance decreases radially outward.  A relatively large mass fraction of the shocked fluid is supported by radial pressure gradients, and only a very small fraction is rotationally supported \citep{Lindner:10}.
Let $r_{\rm rot}$ be defined as the radius at which the radial pressure gradient force and the centrifugal force are equal, $-\rho^{-1} dP/dr= \ell^2/r^3\sim \frac{1}{2}\nabla \Phi$. We will find in the model star that we consider, the black hole strongly dominates the gravitational potential at $r\lesssim 10^4\,\textrm{km}$ at all times (see Section \ref{sec:results} and Figure \ref{fig:profiles}); at these radii, the infalling gas is a negligible perturbation. Thus, here we take that $\nabla\Phi\sim GM_{\rm BH}/r^2$. Given the weak dependence of $\ell$ on radius, since the centrifugal acceleration equals a half of the gravitational acceleration at $r_{\rm rot}$, we have
\beq
\label{eq:r_rot}
r_{\rm rot}\sim 2\frac{\ell^2}{GM_{\rm BH}} .
\eeq  
The radius $r_{\rm rot}$ should not be confused with the circular test particle orbit radius which occurs at $\ell^2/(GM_{\rm BH})$, again treating the potential as Newtonian.
While the fluid is in approximate radial force balance on both sides of $r_{\rm rot}$ as long as the latter radius is contained within $r_{\rm sh}$, the structure of the flow changes character at $r_{\rm rot}$.

At pressure-supported radii, $r\gg r_{\rm rot}$, the contribution of rotation to radial force balance is negligible, and thus a hydrostatic balance can be achieved if the pressure is $P\sim \rho \lambda_P \nabla \Phi$. If the vertical and horizontal pressure scale heights are comparable, i.e., if the flow is thick and radial stratification limits the growing wavelength of the MRI, viscosity can be modeled with \citep{Thompson:05}
\beq
\label{eq:viscosity_Thompson}
\nu\sim \alpha \lambda_P^2 \Omega \ \ \ \ \ \textrm{(pressure support)} ,
\eeq 
where $\lambda_P\equiv|\nabla\ln P|^{-1}$ is the local pressure scale height.  The viscous heating rate is $Q_{\rm visc} = \rho \nu \sigma^2$, where $\sigma=rd\Omega/dr$ is the shear,  With this, the viscous heating rate is
\beq
\label{eq:Q_visc_Thompson}
Q_{\rm visc} \sim \alpha \rho \lambda_P^2 \Omega \left(r\frac{d\Omega}{dr}\right)^2  \ \ \ \ \ \textrm{(pressure support)} .
\eeq

Where the flow is predominantly rotationally supported, $r\ll r_{\rm rot}$, and especially if it collapses into a thin, neutrino-cooled disk, in which the vertical pressure scale height limits the MRI, we would instead have the thin disk value \citep{Shakura:73}
\beq
\label{eq:viscosity_Shakura}
\nu \sim \alpha P/\rho\Omega  \ \ \ \ \ \textrm{(rotational support)} .  
\eeq
We will find that the radii of interest are almost certainly unaffected by neutrino cooling, but the form of viscosity in equation (\ref{eq:viscosity_Shakura}) is still the appropriate one in the region $r<r_{\rm rot}$.  With this,
\beq
\label{eq:Q_visc_Shakura}
Q_{\rm visc} \sim \alpha \frac{P} {\Omega} \left(r\frac{d\Omega}{dr}\right)^2  \ \ \ \ \ \textrm{(rotational support)} .
\eeq

\subsection{A Model for Radial Structure}
\label{sec:radial_structure_model}

With the black hole dominating the gravity, $\nabla\Phi\approx GM_{\rm BH}/r^2$, we model the density, pressure, and specific angular momentum with power-law profiles,
\beq
\label{eq:power_laws}
\rho\propto r^{-\delta},\ \ \ P\propto r^{-\xi}, \ \ \ \ell\propto r^\lambda .
\eeq 
Then, mass continuity $\partial(r^4 v_r \rho)/\partial r=0$ combined with equations (\ref{eq:v_r_viscous}) and (\ref{eq:viscosity_Thompson}) and hydrostatic balance implies that $\xi\approx\delta+1=\lambda+2$.  Note that $\lambda$ can be positive or negative.  With this, for nearly radially-independent specific angular momentum $\lambda\approx 0$, we have that $\xi\approx 2$, so that $P\sim \frac{1}{2} \rho G M_{\rm BH}/r$ and $\nu\sim \frac{1}{4} \alpha \ell$.  Recall that an approximate radial independence of the specific angular momentum is expected given the arguments we have presented in final paragraph of Section \ref{sec:shock_wave} and is seen in rotating, two-dimensional numerical simulations \citep[see, e.g.,][Figure 4]{Lindner:10}. 

At rotationally supported radii, $r\lesssim r_{\rm rot}$, if the disk half-thickness is one half of the radius, $H\sim \frac{1}{2} r$, and vertical pressure balance requires $(H/r)^{-2} (P/\rho)\sim GM_{\rm BH} / r$, it follows that $P \sim \frac{1}{4} \rho GM_{\rm BH}/r$ and from equation (\ref{eq:viscosity_Shakura}) we find that $\nu \sim \frac{1}{4}\alpha GM_{\rm BH}r/\ell$.  Mass continuity and hydrostatic balance now imply  $\xi \approx \delta + 1 \approx 3 - \lambda$.  Note the slightly different numerical coefficient multiplying $\rho GM_{\rm BH}/r$ in the expression for pressure in the rotationally and pressure supported regimes.

In what follows, we adopt, but cannot rigorously justify, the power-law model for the radial dependence of the density and pressure in equation (\ref{eq:power_laws}).  The model is certainly ad hoc, but it does seem to crudely approximate the structure of the solutions that we have obtained with full, time-dependent, hydrodynamical integrations that we present in a separate, companion paper \citep{Lindner:11}, where we find
\beq
\label{eq:power_law_indices}
\xi\approx 2,\ \ \ \delta\approx 1, \ \ \ \lambda\approx 0  \ \ \ \ \ \textrm{(pressure support)} 
\eeq
over a wide range of radii where the shocked fluid is supported by pressure.  In our analytical toy model, unless we explicitly state otherwise, we assume that the power-laws with indices in equation (\ref{eq:power_law_indices}) describe the radial structure of the pressure supported shocked fluid even where viscous quasi-equilibrium has not been reached.  The power-law model allows us to investigate the general properties of collapsar hydrodynamics in the aftermath of the formation of a black hole, and illustrate, as we shall see, the potential for an accretion-powered explosion, but does not grant us the ability to assess the energetics of the explosion with accuracy.

\subsection{The Equation of State in the Innermost Flow}
\label{sec:eos}

To allow us to develop an analytical toy model of the innermost accretion flow, we adopt a simple equation of state.  \citet{Chen:07} showed that the the inner, geometrically thin, neutrino-cooled disk is on the cusp of degeneracy and does not submit itself to reduction to one of the analytically tractable limits \citep[see, also,][]{BisnovatyiKogan:01}.  However, in our companion numerical investigation \citep{Lindner:11}, we find that following the rapid drop in the central accretion rate, the accretion flow no longer cools efficiently and is hot and geometrically thick with the densities and temperatures reaching $\sim 10^8\,\textrm{g}\,\textrm{cm}^{-3}$ and $\sim 2\times10^{10}\,\textrm{K}$, respectively, at the relevant, innermost radii.  Relativistic electrons and pairs dominate the pressure and the adiabatic index remains $\approx 4/3$.   Following a brief transient associated with the initial formation of a rotationally supported torus, degeneracy remains weak, $\mu_e < kT$, within the innermost $\sim 1000\,\textrm{km}$, where $\mu_e$ denotes the electron chemical potential.

To proceed with our highly simplified analysis of the amplitudes the various terms in equation (\ref{eq:energy_conservation}), we will assume that photons and low-density relativistic electrons and pairs dominate the equation of state in the thick disk so that $P\propto T^4$ and $s\propto T^3/\rho$, and will pretend that the impact of nuclear processes on the energetics can be ignored (we lift the latter restriction in Section \ref{sec:nuclear} below). For proton-to-nucleon fractions $Y_e\sim 0.5$ and the relevant range of densities and temperatures, the assumed scalings of the pressure and specific entropy are valid as long as $(30/7\pi^6) (p_{\rm F} c/kT)^6\ll 1$, where $p_{\rm F} = (3\pi^2\rho/2m_p)^{1/3}\hbar$ is the Fermi momentum, or $T\gg 1.4\times10^{10}\,(\rho/10^8\,\textrm{g}\,\textrm{cm}^{-3})^{1/3}\,\textrm{K}$ \citep[see, e.g.,][]{BisnovatyiKogan:01}.  In the hot, geometrically thick, low accretion rate regime, this condition is satisfied in the inner $\sim1000\,\textrm{km}$, where we are about to argue a critical radius should exist within which convection will fail to transport the dissipated energy.  Our adoption of a simple equation of state allows us to construct a crude analytical toy model for the accretion flow; for a more accurate numerical treatment, please see \citet{Lindner:11}.

\subsection{Time Scales}
\label{sec:time_scales}

For reference, here we provide and comment on several time scales characterizing the innermost accretion flow. The time scales can be compared with the one on which 
the energy liberated in central accretion accretion flow energizes the stellar envelope, which in Section \ref{sec:results} we find is rather long, $\sim10\,\textrm{s}$ or longer. We evaluate the characteristic time scales assuming the ad hoc power-law radial scalings in equation (\ref{eq:power_laws}) with the indices given in equation (\ref{eq:power_law_indices}), in the regime in which the flow is supported by pressure.
 In place of the dynamical time scale we quote the time scale for free fall from rest at infinity
\bea
\tau_{\rm FF} &\sim& \frac{r}{v_{\rm FF}}\sim\frac{r^{3/2}}{(2GM_{\rm BH})^{1/2}} \nonumber\\&\sim& 1\,\textrm{ms}\,
\left(\frac{r}{100\,\textrm{km}}\right)^{3/2}\,\left(\frac{M_{\rm BH}}{5\,M_\odot}\right)^{-1/2} .
\eea

The viscous time scale can be estimated by dividing the radius with radial velocity, given in equation (\ref{eq:v_r_viscous}) resulting from angular momentum transport by the viscous torque
\bea
\tau_{\rm visc} &\sim& \frac{r}{v_r} \sim \frac{1}{2} \frac{r^4}{\alpha\,\lambda_P^2\,\ell}
\nonumber\\&\sim& 20\,\textrm{ms}\,\left(\frac{\alpha}{0.1}\right)^{-1}\,\left(\frac{r}{100\,\textrm{km}}\right)^{2}\,\left(\frac{\ell}{10^{17}\,\textrm{cm}^2\,\textrm{s}^{-1}}\right)^{-1} .
\eea
Thus, over the relevant $\sim 10\,\textrm{s}$ time scale, only portion of the shocked, pressure supported material within the inner $\sim 2000\,\textrm{km}$ can accrete viscously.  Since $\tau_{\rm conv}\propto r^{3/2}$ while $\tau_{\rm visc}\propto r^2$ with a larger numerical prefactor, we have $\tau_{\rm conv}\ll \tau_{\rm visc}$ at the larger radii in the shocked fluid that convective cells have reached.  Convection alone will tend to erase gradients in the specific angular momentum profile, consistent with our assumptions that $\lambda\approx 0$, and this is clearly seen in two-dimensional simulations \citep[see, e.g.,][Figure 4]{Lindner:10}.

With the convection we associate the convective eddy radial crossing time scale
\bea
\tau_{\rm conv} &\sim& \frac{\lambda_{\rm conv}}{v_{\rm conv}}\sim\frac{4\,r^{3/2}}{(2GM_{\rm BH})^{1/2}} \nonumber\\&\sim& 4\,\textrm{ms}\,
\left(\frac{r}{100\,\textrm{km}}\right)^{3/2}\,\left(\frac{M_{\rm BH}}{5\,M_\odot}\right)^{-1/2} ,
\eea
which is only somewhat longer than the free fall time scale as a consequence of the steep negative entropy gradient implied by our assumed density profile.

We turn to estimating the time scale for cooling by neutrino emission. At densities $\rho\sim10^7\,\textrm{g}\,\textrm{cm}^{-3}$ (the density dependence is relatively weak) and temperatures $\gtrsim 10^{10}\,\textrm{K}$, the nuclei are almost completely disintegrated and neutrino emission by pair capture onto nucleons (Urca) dominates the cooling.  At lower temperatures, a fraction of the nucleons are in nuclei and neutrino emission by pair annihilation dominates. Ignoring the weak dependence of the free nucleon fraction and of the Urca cooling rate on density, we can write \citep[see, e.g.,][]{Popham:99,DiMatteo:02}
\bea
Q_{\rm cool} &\sim& 10^{25} \,\textrm{ergs}\,\textrm{s}^{-1}\,\textrm{cm}^{-3}\nonumber\\& &\times\cases{ 0.9 \,(T_{\rm K}/10^{10})^6  (\rho_{{\rm g}\,{\rm cm}^{-3}}/10^7), & $T_{\rm K}\gtrsim 10^{10}$ ,\cr
0.5\,(T_{\rm K}/10^{10})^9  , & $T_{\rm K}\lesssim 10^{10}$ .\cr
} 
\eea
With this, the cooling times are 
\bea
& &\tau_{\rm cool} \sim \frac{P}{(\gamma-1) Q_{\rm cool}}\nonumber\\
& & \sim \cases{ 2.2\,\textrm{s}\, (r_{\rm km}/100)^{1/2}\,(\rho_{{\rm g}\,{\rm cm}^{-3}}/10^7)^{-3/2}\,(M_{\rm BH,\odot}/5)^{-1/2}, \cr
0.6\,\textrm{s}\, (r_{\rm km}/100)^{5/4}\,(\rho_{{\rm g}\,{\rm cm}^{-3}}/10^7)^{-5/4}\,(M_{\rm BH,\odot}/5)^{-5/4},\cr
} 
\eea
where the first version applies at temperatures above $\sim 10^{10}\,\textrm{K}$, the second below.  Note for the densities assumed here and verified in Section \ref{sec:results} and Figure \ref{fig:profiles}, we have that $\tau_{\rm cool}\gg \tau_{\rm visc},\, \tau_{\rm conv}$ at all radii, which justifies our neglect of cooling in the treatment of the flow energetics. We provide the time scale for convergence to nuclear statistical equilibrium (NSE) in equation (\ref{eq:tau_NSE}) below.

An additional time scale is the shock crossing time (see Section \ref{sec:results} below)
\beq
\tau_{\rm shock} \sim \frac{r}{v_{\rm sh}} \sim 100\,\textrm{ms}\,\left(\frac{r}{100\,\textrm{km}}\right)\left(\frac{v_{\rm sh}}{10^3\,\textrm{km}\,\textrm{s}^{-1}}\right)^{-1} ,
\eeq
where $v_{\rm sh}$ is the shock velocity.  Comparing $\tau_{\rm conv}$ with $\tau_{\rm shock}$ for the fiducial choice $M_{\rm BH}\sim 5\,M_\odot$, we find that for $v_{\rm sh}\sim 10^4\,\textrm{km}\,\textrm{s}^{-1}$, the convection can keep up with the shock out to $\sim 600\,\textrm{km}$, while for slower shocks $v_{\rm sh}\sim 10^3\,\textrm{km}\,\textrm{s}^{-1}$, convection can keep up until the shock has reached $\sim 6\times10^4\,\textrm{km}$.  Indeed, we will find that that shock slows down considerably after the first $\sim10\,\textrm{s}$.  This justifies our assumption that convection is fully developed in the shocked stellar envelope and can transport energy from the innermost accretion flow toward the the shock wave in the course of its traversal of the progenitor's interior.  The shock, however, must eventually decouple and re-accelerate in the outer atmosphere.

\subsection{The Failure of Convection at Small Radii}
\label{sec:convection}

We will attempt to compare the relative amplitudes of the advection term $Q_{\rm adv}\equiv v_r\rho Tds/dr$, the convection term $Q_{\rm conv}\equiv \vec{\nabla}\cdot\vec{F}_{\rm conv}=r^{-2} d(r^2F_{\rm conv})/dr$, and the viscous heating term $Q_{\rm visc}$ in equation (\ref{eq:energy_conservation}).  The detailed form of these terms depends on the equation of state; we work assuming that $P\propto T^4$ and $s\propto T^3/\rho$ (see Section \ref{sec:eos}). 
In the pressure-supported regime, with hydrostatic balance implying $\xi\approx \delta+1$, we find $Q_{\rm adv}\propto r^{-2-\xi+\lambda}$, while the convective term is $Q_{\rm conv}\propto r^{-3/2-\xi}$, thus the ratio of the advection to the convective term is $Q_{\rm adv}/Q_{\rm conv}\propto r^{-1/2+\lambda}$, which for rotation laws $\lambda<1/2$ increases inward. This opens the possibility for the presence of an advection-dominated accretion flow (ADAF) at the very smallest radii; at larger radii, we have a convection-dominated accretion flow (CDAF; see, e.g., \citealt{Stone:99,Igumenshchev:00b,Blandford:04}). In the rotationally-supported regime, $Q_{\rm adv}\propto r^{-1-\xi-\lambda}$ and $Q_{\rm adv}/Q_{\rm conv}\propto r^{1/2-\lambda}$, implying that, to the extent that convective transport in the rotationally-supported regime (see Section \ref{sec:shock_wave}) can be modeled with MLT---this assumption can only be tested and calibrated with multidimensional hydrodynamic simulations---the smallest radii favor a CDAF. Thus for estimating the luminosity carried by the shocked fluid toward the expanding shock wave, it is key to pin down the radius $r_{\rm ADAF}$ of the ADAF-CDAF transition and its relation to the the radius $r_{\rm rot}$ separating the inner rotationally supported region from the outer, pressure supported region.  The radii $\sim r_{\rm rot}$ seem to be the most susceptible to the appearance of an ADAF.

The ADAF regime can occur at radii smaller than a transition radius $r_{\rm ADAF}$ where the spherically-averaged advection fluxes in equation (\ref{eq:total_energy_conservation}), in brackets, exceed the convection fluxes $F_{\rm conv}$ and $F_{\rm mix}$. This, in general, differs from the criterion requiring that $(Q_{\rm adv},Q_{\rm conv})>Q_{\rm visc}$ in the equatorial region, which has been employed elsewhere \citep[e.g.,][]{Kohri:05}, since we distinguish spherically-averaged advection and spherically-averaged convection.  In what follows we will conservatively assume that an ADAF is present and
\beq
r_{\rm ADAF}\gtrsim r_{\rm rot} 
\eeq
so that at $r<r_{\rm ADAF}$, the bulk of the dissipated energy travels inward.  If ADAF is absent and CDAF extends to the innermost radii, then $r_{\rm ADAF}$ in the forthcoming development should be replaced by $r_{\rm ISCO}$; in this case, the convection must carry most of the energy dissipated according to equation (\ref{eq:Q_visc_Shakura}).

Comparing the negative (radially inward) radial advection luminosity 
\beq
\label{eq:luminosity_advection}
L_{\rm adv}=4\pi r^2 v_r \frac{\gamma P}{\gamma-1}
\eeq
to the positive (radially outward) accretion luminosity 
\beq
\label{eq:luminosity_accretion}
L_{\rm acc} =4\pi r^2 v_r\rho \Phi 
\eeq
in the pressure-supported regime we find that $L_{\rm adv} \propto r^{-\xi+\lambda+1}$ while $L_{\rm acc}\propto r^{-1}$. Again, since $\xi\approx\delta+1=\lambda+2$  for $r>r_{\rm rot}$, we find that $L_{\rm adv} = -2 L_{\rm acc} \propto r^{-1}$.  On the other hand, the convection luminosity 
\beq
\label{eq:luminosity_convection}
L_{\rm conv}=4\pi r^2 F_{\rm conv}
\eeq 
is $L_{\rm conv}\propto r^{-\xi+3/2}\propto r^{-\lambda-1/2}$.  In this regime, the energy flux due to kinetic energy advection is minus one-half of the viscous torque; their sum should be relatively small at the radii where $L_{\rm conv}$ dominates energy transport. With $\lambda<\frac{1}{2}$, therefore, there is a radius at which the positive convection luminosity $L_{\rm conv}$ cannot compete with the inward advection luminosity reduced by the outward accretion luminosity $L_{\rm adv}+L_{\rm acc}$ as long as pressure support holds. Note that the viscous luminosity 
\beq
L_{\rm visc}=-4\pi r^2\rho\nu\ell \frac{d\Omega}{dr}
\eeq
is relatively small compared to $L_{\rm adv}$ and $L_{\rm acc}$.  

\subsection{Estimates of the Luminosity}
\label{sec:luminosity}

The luminosity that is transported outward from the smallest radii through the convective shocked stellar envelope is challenging to estimate because of the non-self-similar nature of the accretion flow. In this section, we consider energy transport by hydrodynamic processes: advection, viscous stress, and convection; we recall that cooling by neutrino emission is inefficient and defer addressing of the role of nuclear compositional transformation in the transport of energy to the following section.  To offset theoretical uncertainties, we attempt to place multiple constraints on the luminosity.   If the specific angular momentum in the shocked region is nearly independent of radius, $|\lambda|\ll \frac{1}{2}$, mass continuity at the radii where $\partial\ell/\partial t\sim 0$ implies $\delta\approx 1$.   As indicated in Section \ref{sec:radial_structure_model}, we adopt an ansatz whereby $\xi= 2$, $\delta=1$, and $\lambda\approx 0$ throughout the pressure-supported section of the shocked region, even the radii where the flow has not reached viscous quasi-equilibrium, so that $\rho(r)=\rho(r_{\rm ADAF}) r_{\rm ADAF}/r$.  Since the sum of advection, accretion, and convection luminosities, $L_{\rm adv}+L_{\rm acc}+L_{\rm conv}$, must be independent of radius in a quasi-steady state, the pressure profile at the radii where convection does not dominate energy transport should rise inward less steeply than radial hydrostatic balance implies; the latter is possible, of course, because of the increasing role of rotation at small radii.  

We hypothesize that $L_{\rm adv}(r_{\rm rot}) +L_{\rm acc}(r_{\rm rot}) \sim 0$, which is indeed satisfied if $P(r\lesssim r_{\rm rot}) \sim \frac{1}{4} \rho GM_{\rm BH}/r$ as we suggested above; thus, the net luminosity flowing through this radius, ignoring nuclear and neutrino contributions, is simply the convection luminosity $L=L_{\rm conv}(r_{\rm rot})$.  With this, we have that $L=(L_{\rm adv}+L_{\rm acc}+L_{\rm conv} )(r_{\rm rot}) \sim L_1$, where in the toy model with radiation-like equation of state (Section \ref{sec:eos}), from equations (\ref{eq:convective_flux}) and (\ref{eq:r_rot}),
\beq 
\label{eq:luminosity_1}
L_1\equiv L_{\rm conv}(r_{\rm rot})\sim \pi GM_{\rm BH} \ell \rho(r_{\rm rot}) \left[\frac{\lambda_{\rm conv}(r_{\rm rot})}{r_{\rm rot}}\right]^2 .
\eeq
Note that $L_1$ does not depend on the viscosity parameter $\alpha$.

The ADAF-CDAF transition radius $r_{\rm ADAF}$ can be defined as the radius at which $L_{\rm adv}=-L_{\rm conv}$. From this and equations (\ref{eq:convective_flux}), (\ref{eq:luminosity_advection}), and (\ref{eq:luminosity_convection}) we obtain 
\beq
\label{eq:r_ADAF}
r_{\rm ADAF}\sim 32\alpha^2\frac{ \ell^2}{ GM_{\rm BH}} \left[\frac{\lambda_{\rm conv}(r_{\rm ADAF})}{r_{\rm ADAF}}\right]^{-4} ,
\eeq 
which is larger than $r_{\rm rot}$ when $\lambda_{\rm conv}(r_{\rm ADAF})< 2\alpha^{1/2} r_{\rm ADAF}$. The advection radius $r_{\rm ADAF}$ is larger by a dimensionless factor of the form $\sim 16\, [\lambda_{\rm conv}(r_{\rm ADAF})/r_{\rm ADAF}]^{-2}$ than the critical radius, discussed by \citet{Abramowicz:02}, at which the accretion velocity $|v_r|$ exceeds the velocity of convective cells $v_{\rm conv}$.

Combining equation (\ref{eq:r_ADAF}) with equation (\ref{eq:luminosity_accretion}) we have $L=(L_{\rm adv}+L_{\rm acc}+L_{\rm conv} )(r_{\rm ADAF})\sim L_2$ where 
\beq
\label{eq:luminosity_2}
L_2\equiv L_{\rm acc}(r_{\rm ADAF}) \sim 2\pi \alpha GM_{\rm BH} \ell \rho(r_{\rm ADAF}) .  
\eeq
Since convection can carry the luminosity $L$ to the accretion shock, the energy available to power an explosion is also sensitive to $\ell$, $\lambda_{\rm conv}$, and $\alpha$; it will be also sensitive to the nuclear processes that we have neglected in this toy model thus far.  Of course, if $\alpha$ is particularly small, neutrino cooling and the settling of the flow into a thin disk may start competing with advection at the smallest radii.  

An additional lower limit on $r_{\rm ADAF}$ can be placed by noting that this is the smallest radius such that convection can transport all the energy dissipated exterior to the radius, 
\beq
\int_{r_{\rm ADAF}}^{r}  Q_{\rm visc}  4\pi r^2 dr \leq L_{\rm conv}(r)\  \textrm{  for any  }\ r>r_{\rm ADAF}
\eeq 
which, with equations(\ref{eq:convective_flux}),  (\ref{eq:Q_visc_Thompson}), and (\ref{eq:luminosity_convection}) yields the condition 
\beq
\label{eq:r_ADAF_limit}
r_{\rm ADAF}\geq 2\alpha^{2/3}\frac{\ell^2}{GM_{\rm BH}} \left[\frac{\lambda_{\rm conv}(r_{\rm ADAF})}{r_{\rm ADAF}}\right]^{-4/3} .
\eeq
This differs from our previous estimate only by a numerical constant of the order of unity and should be interpreted as only a lower limit on the ADAF-CDAF transition because of the possibility that advection or nuclear energy transfer through disintegration, recombination, and compositional mixing is still able to carry inward some of the energy dissipated outside this radius.

The density at the rotation and the advection radius can be related to the immediate post-shock density via $\rho(r_{\rm rot}) r_{\rm rot} \sim \rho(r_{\rm ADAF}) r_{\rm ADAF}\sim\rho(r_{\rm sh}) r_{\rm sh}$.  With this, we have that, for $\alpha\sim 0.1$, the two different luminosity estimates are comparable, $L_2/L_1\sim (8\alpha)^{-1} [\lambda_{\rm conv}(r_{\rm ADAF})/r_{\rm ADAF}]^2 \sim {\cal O}(1)$, which lends support to the consistency of our two methods to estimate the luminosity carried by the convective envelope.  We can thus combine equations (\ref{eq:luminosity_1}) and (\ref{eq:luminosity_2}) to write 
\bea
\label{eq:luminosity_cases}
L &\sim& \left[1,\frac{1}{8\alpha}\left(\frac{\lambda_{\rm conv}}{r}\right)^2_{r_{\rm ADAF}}\right]\frac{\pi}{2}\frac{(GM_{\rm BH})^2}{\ell} \nonumber\\ & & \times\left(\frac{\lambda_{\rm conv}}{r}\right)^2_{r_{\rm ADAF}}r_{\rm sh}\rho(r_{\rm sh}) ,
\eea
where the dimensionless coefficients in brackets correspond to $L\sim L_1$ and $L\sim L_2$, respectively. 

If the gravitational potential is approximately Keplerian, hydrostatic balance in the pressure-supported outer parts of the post-shock region requires that $P\propto r^{-\xi}$ with $\xi=2$, and thus, most of the mass and energy resides close to the shock.  In a medium in which relativistic, low-density electrons and pairs dominate the equation of state, $4\pi r^2 F_{\rm conv}\propto r^{-\xi+3/2}$, hence $(\delta,\xi)=(1,2)$ does not ensure a radius-independent convective luminosity.  In reality, some of the convective luminosity is converted into bulk motion with $v_r>0$, and weakly relativistic and degenerate electrons and pairs will dominate the equation of state; these effects will reconcile the radial momentum equation for the nonrotating outer region, $v_r dv_r/dr=-\nabla\Phi-\rho^{-1}dP/dr$,
with the convective flux conservation $d(r^2F_{\rm conv})/dr=0$ to yield the correct, and likely radially dependent, indices $\delta$ and $\xi$.  We allow our toy model to depart from local self-consistency; instead, in Section \ref{sec:conservation} below, we require global mass and energy conservation.  Beforehand, however, we must address the possibility that nuclear compositional transformation contributes to radial energy transport.

\subsection{Nuclear Disintegration and Recombination}
\label{sec:nuclear}

Our requirement that neutrino cooling be inefficient at $r_{\rm ISCO}$ can clearly be relaxed to require that it be inefficient at $r_{\rm ADAF}$; this is generally the case for relatively large $\alpha$ and central accretion rates $\dot M\ll 0.1\,M_\odot\,\textrm{s}^{-1}$.  Relativistic effects are also weak outside $r_{\rm ADAF}$, especially when the black hole rotates rapidly. A more significant concern is the possibility that nuclear composition changes as material passes the shock and arrives at $r_{\rm ADAF}$.  At densities expected in the vicinity of $r_{\rm ADAF}$, which are $\sim 10^6-10^8\,\textrm{g}\,\textrm{cm}^{-3}$, and on time scales $\sim 0.1-10\,\textrm{s}$, the principal nucleosynthetic products reach nuclear statistical equilibrium (NSE) conditions at temperatures $T\gtrsim 4\times10^9\,\textrm{K}$. 
For proton-to-nucleon fractions $Y_e\sim 0.5$, the iron-group-to-helium transition takes place at $T\sim (5-7)\times10^9\,\textrm{K}$, and the helium-to-free nucleon transition takes place at $T> (7-10)\times10^9\,\textrm{K}$.  These temperatures can be compared to an estimate of the temperature in our model at pressure supported radii, $r>r_{\rm rot}$, that reads
\bea
T&\sim& 1.3\times10^{10}\,\textrm{K} \,\beta_{\rm rad,1/3}\left(\frac{GM_{\rm BH}r}{\ell^2}\right)^{-1/4}\left(\frac{M_{\rm BH}}{5\,M_\odot}\right)^{1/4}
\nonumber\\& &\times 
\left(
\frac{\rho}{10^7\,\textrm{g}\,\textrm{cm}^{-3}}\right)^{1/4}\left(\frac{\ell}{10^{17}\,\textrm{cm}^2\,\textrm{s}^{-1}}\right)^{-1/2} ,
\eea
where $\beta_{\rm rad,1/3}$ is the fraction of pressure in relativistic species to total pressure in units of one third.  The temperature of the accretion flow clearly straddles the temperatures at which compositional transitions occur.

If the helium-to-nucleon transition occurs at radii $r_{\rm dis}$ comparable to or smaller than $r_{\rm ADAF}$, which can be the case during a period following shock formation, then the net energy flux through $r_{\rm ADAF}$ should be augmented with
\bea
\label{eq:nuclear_luminosity}
L_{\rm nuc}&\sim& \left\{4\pi r^2 \rho \left[v_r(\epsilon_{\rm nuc}-\epsilon_{\rm nuc}^\infty) - \frac{1}{3}\chi_{\rm mix} v_{\rm conv} \lambda_{\rm conv} \frac{d\epsilon_{\rm nuc}}{dr} \right]\right\}_{r_{\rm ADAF}} ,
\eea
where $\epsilon_{\rm nuc}^\infty\approx -8\,\textrm{MeV}/m_p$ is the specific (negative)  pre-shock nuclear binding energy of the material at the advection radius.   The second term in brackets in equation (\ref{eq:nuclear_luminosity}) represents the outward energy transport arising from the convective compositional mixing.  If, with the help of equations  (\ref{eq:v_r_viscous}) and (\ref{eq:viscosity_Thompson}), we identify
\beq
\dot M = -4\pi r^2 \rho v_r \sim 2\pi \alpha \ell r_{\rm sh} \rho(r_{\rm sh}) 
\eeq 
with the rate  with which shocked material accretes onto the black hole, and further identify
\beq
\dot M_{\rm conv} = 4\pi r^2 \rho v_{\rm conv}
\eeq 
with the rate with which convective cells transport mass radially, then we can rewrite equation (\ref{eq:nuclear_luminosity}) as
\bea
\label{eq:nuclear_luminosity_M_dot}
L_{\rm nuc}&\sim& \left[(\epsilon_{\rm nuc}^\infty-\epsilon_{\rm nuc}) \dot M - \frac{1}{3}\chi_{\rm mix} \frac{\lambda_{\rm conv}}{r} \frac{d\epsilon_{\rm nuc}}{dr} \dot M_{\rm conv}\right]_{r_{\rm ADAF}} .
\eea
It is typical of CDAFs that $\dot M_{\rm conv}\gtrsim \dot M$, and thus, depending on the relative location of $r_{\rm dis}$ and $r_{\rm ADAF}$, the mixing term may reduce the parameter space in which $L_{\rm nuc}<0$ and could potentially even lead to $L_{\rm nuc}>0$.

\subsection{Summary on Advection and Convection}
\label{sec:summary_advection_convection}

Our goal has been to estimate the rate with which energy is carried toward larger radii of the shocked stellar atmosphere where the flow has the potential to become ``dynamical'' and engender an explosion. If given such an estimate of the luminosity of the inner accretion flow, we are in position to investigate the global, time-dependent hydrodynamics of the star by treating the central luminosity as a source and by requiring global energy conservation. This is the subject of the following section.   To attempt to estimate the luminosity, we have examined radial energy transport in the inner accretion flow which occurs in an approximate quasi-steady-state so that the time derivative term in the spherically-averaged energy transport equation can be neglected and radial force balance is a good approximation. Unfortunately, exact analytical treatment is complicated by the radially increasing dominance of pressure support over rotational support in our model.  For this reason, we have steered away from attempting to identify a formal solution to the eigenvalue problem defined by the conservation of mass, angular momentum, and energy in a flow allowing for advection as well as convection \citep[see, e.g.,][who solve the eigenvalue problem under simplifying assumptions that do not apply in the present context]{Abramowicz:02,Lu:04}.  Instead, we have adopted an intuitive approach in which we make educated guesses about the radial scaling of the fluid variables.  

Our intuitive approach has allowed us to conclude that while the dominant energy transport term at large radii is convection, at small radii, convection should not be able to compete with advection, implying that the flow transitions from a CDAF to an ADAF at some critical radius.  This critical radius will generally depend on the viscous stress-to-pressure ratio $\alpha$, the efficiency of the convection (which we parameterize in terms of the mixing length), and the angular momentum of the accreting fluid, but in our approximate treatment, we are not able to pin it down with absolute certainly; instead, we provide an approximation in equation (\ref{eq:r_ADAF}) and a lower limit in equation (\ref{eq:r_ADAF_limit}).  Only the energy dissipated outside the critical radius will flow outward and contribute to the luminosity of the central source.  We have obtained two slightly different estimates of the luminosity, which we summarize in equation (\ref{eq:luminosity_cases}), and have also estimated the impact of energy transport by nuclear disintegration and recombination in equation (\ref{eq:nuclear_luminosity_M_dot}).  In the following section, we incorporate the luminosity estimate in dynamical model of the envelope surrounding the inner, rotationally-supported accretion flow.

\section{Mass and Energy Conservation}
\label{sec:conservation}

We proceed to study the global dynamics of the stellar envelope outside of the innermost, rotationally-supported accretion flow. 
A fraction of the envelope has passed the accretion shock; the kinematics of the shock, as we shall see, is the principal determinant of the conditions in the innermost flow, while the luminosity transported outward from the innermost flow is the driver of the shock's dynamics.  Keeping in mind this interdependence, we model the density and radial velocity structure of the fluid flow inside and outside the shock radius and require mass and energy conservation.  Normally, mass, momentum, and energy conservation are imposed in the form of jump conditions applied to the shock transition itself, but this approach generally leads to violation of global mass and energy conservation, unless one solves the full, time- and radius-dependent hydrodynamic transport equations.  For the purpose of analytic transparency, we opt for a simpler approach in which we adopt a model density and velocity field, and then require global mass and energy conservation.  This may violate local conservation at the shock.

If at mass coordinates just exterior to those first passing through a nascent accretion shock, the stellar pre-supernova density profile is approximately $\rho_{\star,t=0}\sim M_\star/4\pi r_\star r^2$, where $M_\star$ the stellar mass enclosed within some radius $r_\star$, then the subsequent flow of material into the shock may resemble the self-similar collapse of nearly-hydrostatic isothermal spheres of \citet{Shu:77}.  This solution is characterized by a critical point (a rarefaction front) that recedes into the pre-collapse envelope at the speed of sound.  The critical point starts traversing the star outward at $t=0$, and thus it is well ahead of the shock front, which starts traveling after tens of seconds, even if the shock travels supersonically.  The density profile of the collapsing envelope in the self-similar solution steepens from $\rho_\star\propto r^{-1}$ in the immediate vicinity of the critical point to $\rho_\star \propto r^{-3/2}$ at radii $r\ll r_{\rm crit}\sim (5-10)\times10^{4}\,\textrm{km}$, where the critical point traverses the star at the sound speed $dr_{\rm crit}/dt = (\gamma_\star G M_\star /2r_\star)^{1/2}$, where $\gamma_\star$ is the adiabatic index of the stellar envelope.  We adopt $\rho_\star\propto r^{-3/2}$ for the infalling envelope. The fluid velocity can be approximated via $v_\star\sim (1-r/r_{\rm crit})v_{\rm ff} $, where $v_{\rm ff}(r,t)=-[GM(r,t)/r]^{1/2}$ is the free fall velocity.

Let $M_\star$ be the stellar mass enclosed within some radius $r_\star$.  Then we model the stellar density profile via
\beq
\label{eq:density_regions}
\rho(r) = \cases{ 
M_{\rm sh}/(2\pi r_{\rm sh}^2 r), & $r < r_{\rm sh}$, \cr
M_\star/(4\pi r_\star r_{\rm crit}^{1/2} r^{3/2}), & $r_{\rm sh} < r < r_{\rm crit}$, \cr
M_\star/(4\pi r_\star r^2) , & $r_{\rm crit} < r < r_\star$, 
}
\eeq
where 
\beq
M_{\rm sh}=2\pi r_{\rm sh}^3\rho(r_{\rm sh}) 
\eeq
is the mass of the shocked fluid.  The equation of mass continuity in the post-shock fluid, $\partial\rho/\partial t+r^{-2} \partial(r^2 v_r \rho)/\partial r=0$, implies, for the outer region $r\sim r_{\rm sh}$ in which viscous accretion can be ignored
\beq
v_r \sim r \frac{d}{dt} \ln \frac{r_{\rm sh}}{M_{\rm sh}^{1/2}} ,
\eeq
and with this we model the radial velocity profile via
\beq
\label{eq:velocity_regions}
v_r(r) = \cases{ 
r d\ln(r_{\rm sh}/M_{\rm sh}^{1/2})/dt, & $r < r_{\rm sh}$, \cr
-(1-r/r_{\rm crit}) [GM(r,t)/r]^{1/2} , & $r_{\rm sh} < r < r_{\rm crit}$, \cr
0, & $r_{\rm crit} < r < r_\star$ ,
}
\eeq
where $M(r,t)=M_{\rm BH}+\int_0^r 4\pi \rho r^2 dr$.  Finally, in the outer parts of the shocked region, $r\sim r_{\rm sh}$, the momentum of the shocked fluid contributes negligibly to radial pressure balance, and thus the shocked region is quasi-hydrostatic.  We model the pressure profile via
\beq
\label{eq:pressure_regions}
P(r) = \cases{ 
\frac{1}{2}\rho r \nabla \Phi, & $r < r_{\rm sh}$, \cr
\frac{1}{2}[\rho r \nabla \Phi]^{r_{\rm crit}}_{r_\star} (r/r_{\rm crit})^{-\gamma_\star}, & $r_{\rm sh} < r < r_{\rm crit}$, \cr
\frac{1}{2}[\rho r \nabla \Phi]^r_{r_\star}, & $r_{\rm crit} < r < r_\star$, 
}
\eeq
where, again, $\Phi$ is the total gravitational of the black hole and the mass distribution in equation (\ref{eq:density_regions}).

Mass conservation requires that 
\beq
\label{eq:mass_conservation}
\frac{d}{dt}\left(M_{\rm BH}+M_{\rm sh}  + \int_{r_{\rm sh}}^{r_\star} 4\pi r^2 \rho dr \right)=0 .
\eeq
The mass of the black hole at the moment of shock formation is 
\beq
M_{\rm BH}(t_{\rm sh})=M_{\rm BH,0}+\frac{1}{3} \frac{r_{\rm crit}(t_{\rm sh})}{r_\star} M_\star, 
\eeq
where $M_{\rm BH,0}$ is the initial black hole mass immediately following core collapse in excess of the mass available in the core from the inward-extrapolated profile $\rho_\star\propto r^{-2}$.
Since the accretion rate into the black hole just prior to shock formation $\dot M(t<t_{\rm sh}) \sim \frac{1}{3} (M_\star/r_\star) dr_{\rm crit}/dt$ is similar to the accretion rate into the shock after shock formation, and the black hole accretion rate experiences a drop as the shock starts traveling outward, we expect $dM_{\rm sh}/dt\gg dM_{\rm BH}/dt$.  

Pretending that $r_\star\sim 10^5\,\textrm{km}$ is the true edge of the star and that efficient neutrino cooling, if anywhere, occurs only at $r<r_{\rm ADAF}$, global energy conservation can be written as 
\bea
\label{eq:global_energy_conservation}
L + L_{\rm nuc}&=& \frac{d}{dt} (U+K+W+E_{\rm nuc}) 
\nonumber\\
-
& & 
\int_{r_{\rm ADAF}}^{r_\star} \frac{d}{dt}\left(-\frac{GM_{\rm BH}}{r}\right)4\pi\rho r^2 dr ,
\eea
where $L$  is the net energy flow rate at $r_{\rm ADAF}$, while $U$, $K$, $W$, and $E_{\rm nuc}$, are the total internal, kinetic, gravitational potential, and nuclear energies in the annulus $r_{\rm ADAF}<r<r_\star$.  The contribution of nuclear composition change to the energy flux through the advection radius, $L_{\rm nuc}$, is significant only when the helium-to-free nucleon disintegration occurs at radii $\gtrsim r_{\rm ADAF}$.  After the shock has expanded far beyond the disintegration radius, we can assume that the mass in free nucleons evolves very slowly in time, and thus, $dE_{\rm nuc}/dt\approx 0$.  

The total internal energy is the sum over the shocked and unshocked regions
\beq
\label{eq:energy_inside_shock}
U \approx \int_{r_{\rm ADAF}}^{r_{\rm sh}} \frac{P}{\gamma-1} 4\pi r^2 dr +\int_{r_{\rm sh}}^{r_\star} \frac{P}{\gamma_\star-1} 4\pi r^2 dr  ,
\eeq
where the pressure is calculated following the model in equation (\ref{eq:pressure_regions}).  
The kinetic energy is calculated via
\beq
K= \int_{r_{\rm ADAF}}^{r_{\rm crit}} \frac{1}{2} \rho v_r^2 4\pi r^2 dr .
\eeq
The total gravitational energy $W$ of the density in equation (\ref{eq:density_regions}) in the presence of self gravity and the gravity of the black hole is straightforward to calculate,
\beq
W = \int_{r_{\rm ADAF}}^{r_\star} \left(-\frac{GM_{\rm BH}}{r} + \frac{1}{2}\Phi_\rho \right) \rho \,4\pi r^2 dr ,
\eeq
where the gravitational potential of the fluid outside of the black hole is related to the density via $4\pi G\rho= \nabla^2 \Phi_\rho$.

\begin{figure}
\begin{center}
\includegraphics[width=3.5in]{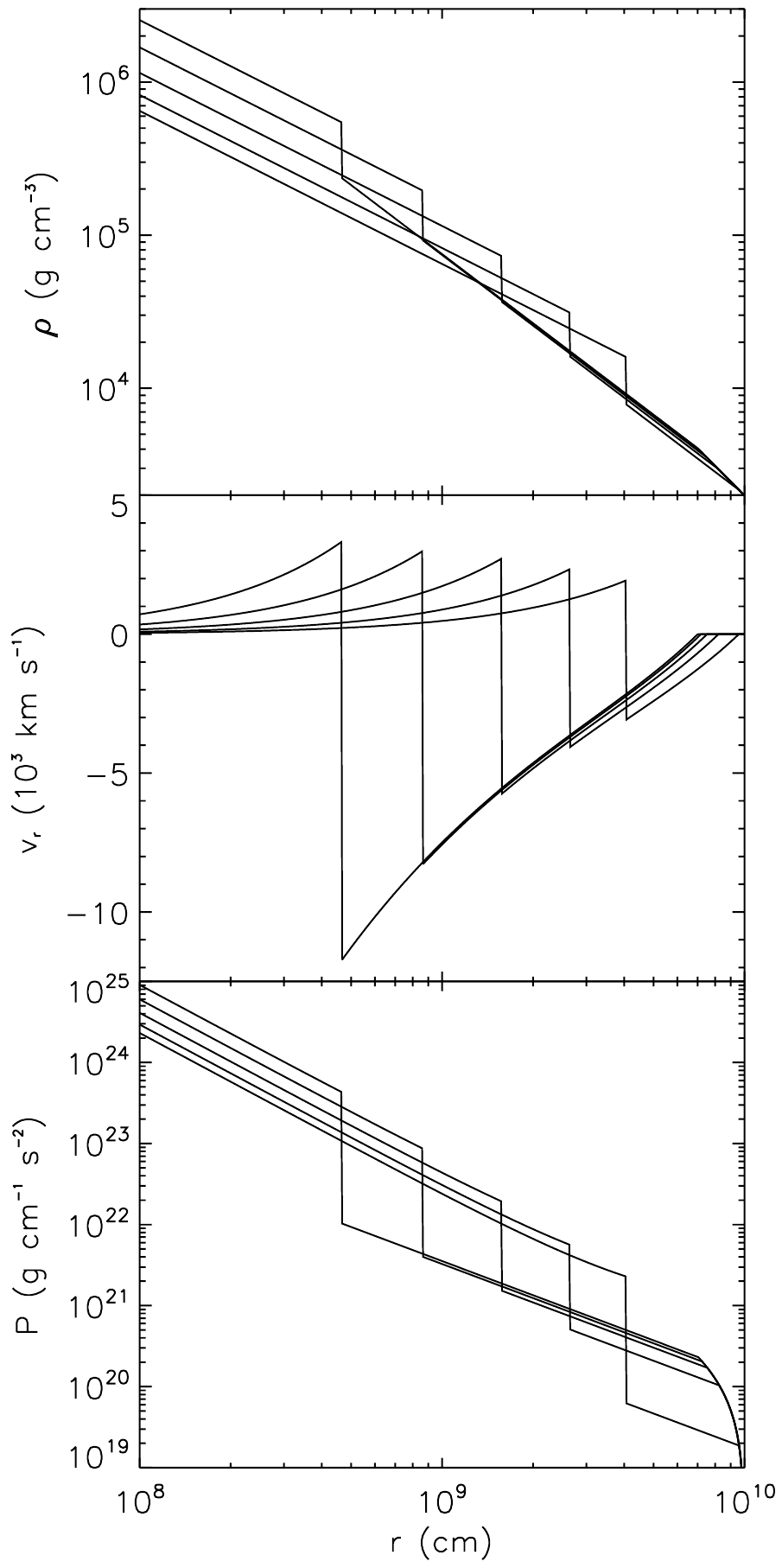}
\end{center}
\caption{The density $\rho$, radial velocity $v_r$, and pressure $P$ of the toy model as given by equations (\ref{eq:density_regions}), (\ref{eq:velocity_regions}), and (\ref{eq:pressure_regions}), for $M_\star=12.5\,M_\odot$, $r_\star=10^{10}\,\textrm{cm}$, $\gamma_\star=1.4$, $M_{\rm BH,0}=2.5\,M_\odot$, $\ell=10^{17}\,\textrm{cm}^2\,\textrm{s}^{-1}$, $\alpha=0.1$, $\lambda_{\rm conv}/r=0.75$, $L_{\rm nuc}=0$, $t_{\rm sh}=20\,\textrm{s}$, and $t-t_{\rm sh}=(0.5,1,2,4,8)\,\textrm{s}$; the convective luminosity was calculated from equation (\ref{eq:luminosity_2}). The radial velocity does not take into account the negative radial velocity resulting from viscous angular momentum transport. The density jump at the shock is smaller than it should be because our adopted density profile for the infalling envelope, $\rho\propto r^{-3/2}$ for $r_{\rm sh}<r<r_{\rm crit}$ is too steep at radii just smaller than $r_{\rm crit}$, where $\rho\propto r^{-1}$.}
\label{fig:profiles}
\end{figure}

\section{Shock Dynamics}
\label{sec:results}

Equations (\ref{eq:mass_conservation}) and (\ref{eq:global_energy_conservation}) can be 
construed as constraints relating the shock velocity to the shock radius and time, $v_{\rm sh}(r_{\rm sh},t)$.  The equation $dr_{\rm sh}/dt=v_{\rm sh}$ can then be integrated to solve for $r_{\rm sh}(t)$.  Since we expect that the shock velocity varies slowly in time, we do not carry out the formal integration and instead approximate 
\beq
\label{eq:r_sh_approx}
r_{\rm sh}\sim  (t-t_{\rm sh})  v_{\rm sh} ,
\eeq 
where, as before, $t_{\rm sh}$ denotes the shock formation time.  This allows us to estimate $v_{\rm sh}(t)$ and $r_{\rm sh}(t)$.  In Figure \ref{fig:profiles}, we show a typical time evolution of the density, radial velocity, and pressure profile for a fiducial stellar model with $M_\star=12.5\,M_\odot$, $r_\star=10^{10}\,\textrm{cm}$, $\gamma_\star=1.4$, $M_{\rm BH,0}=2.5\,M_\odot$, and $\ell=10^{17}\,\textrm{cm}^2\,\textrm{s}^{-1}$. This model approximates the density profile of the fully mixed pre-supernova Wolf-Rayet model 16TI of \citet{Woosley:06a} in the radial range $2\times10^8\,\textrm{cm}\lesssim r \lesssim 5\times10^9\,\textrm{cm}$.  At radii $r\gtrsim \frac{1}{2} r_\star$, our toy model overestimates the stellar envelope density, which declines increasingly steeply with radius, and thus it underestimates the shock velocity after the first $\sim 10\,\textrm{s}$.  Note that the black hole dominates the mass enclosed for $r\ll 4\times 10^4\,\textrm{km}$ at all times, which justifies our neglect of the infalling mass in the calculation of the force balance in Section \ref{sec:support}.

The density jump at the shock in Figure \ref{fig:profiles} is substantially smaller than it should be; e.g., in the strong-shock limit---which is not always reached here---the density jumps sevenfold for $\gamma=\frac{4}{3}$.  The anomaly seems to be an artifact of our assumption that the density in the region $r_{\rm sh}< r< r_{\rm crit}$ is a pure power law $\rho\propto r^{-3/2}$.  In reality, as illustrated in the \citet{Shu:77} solution for the self-similar collapse of isothermal spheres, the density profile is less steep, $\rho\propto r^{-1}$, just inside $r_{\rm crit}$, and thus our immediate pre-shock density is an overestimate.  Since the immediate post-shock density is calculated independently from global mass conservation, the ratio of the two densities as seen in Figure \ref{fig:profiles} is also an overestimate.  In spite of this, the model does conserve mass and energy globally.

\begin{figure}
\begin{center}
\includegraphics[width=3.5in]{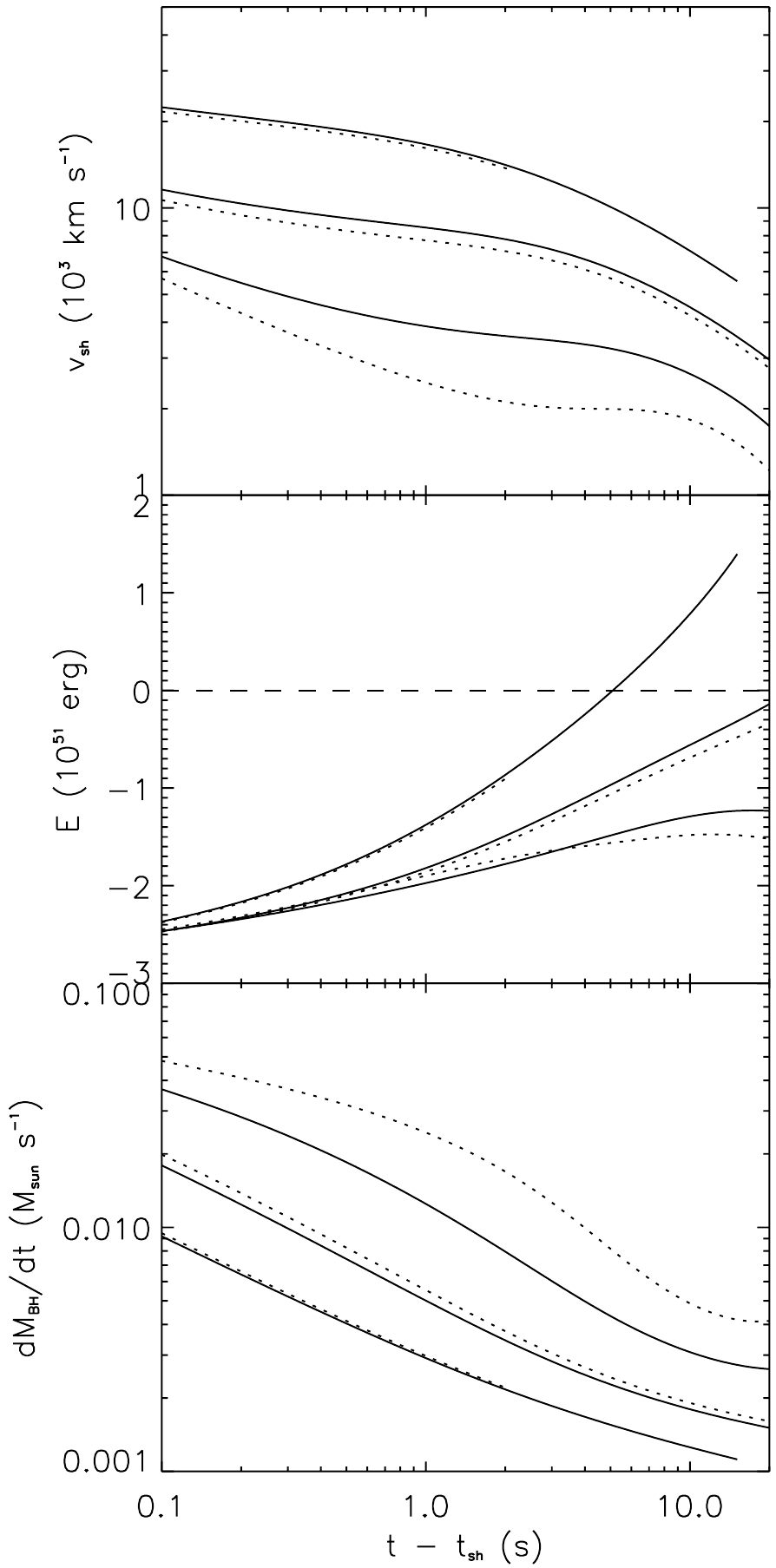}
\end{center}
\caption{The shock velocity $v_{\rm sh}$, the total energy of the fluid outside the black hole $E$, and the accretion rate onto the black hole $\dot M\equiv dM_{\rm BH}/dt$ as a function of time from shock formation, $t-t_{\rm sh}$, for the fiducial model with $M_\star=12.5\,M_\odot$, $r_\star=10^{10}\,\textrm{cm}$, $\gamma_\star=1.4$, $M_{\rm BH,0}=2.5\,M_\odot$, $\ell=10^{17}\,\textrm{cm}^2\,\textrm{s}^{-1}$, $\alpha=0.1$, and $t_{\rm sh}=20\,\textrm{s}$; the convective luminosity was calculated from equation (\ref{eq:luminosity_2}).  The solid curves are ignoring nuclear disintegration, $L_{\rm nuc}=0$, for $\lambda_{\rm conv}/r=(0.5,0.75,1.0)$; larger convective mixing length give faster shocks, more energy deposition, and lower accretion rates.  The dotted lines are the same but with maximum energy loss due to nuclear disintegration, $L_{\rm nuc}=-8\,\textrm{MeV}\,\dot M/m_p$.  The pre-shock accretion rate, not shown in the figure, is $\dot M(t<t_{\rm sh}) \approx 0.14\,M_\odot\,\textrm{s}^{-1}$.}
\label{fig:evolution}
\end{figure}

For the viscous stress-to-pressure ratio, we adopt $\alpha=0.1$ and in Figure \ref{fig:evolution}, we show the evolution of the shock velocity, the total energy of the stellar envelope, and the rate with which the envelope is accreting onto the black hole for three values of the convective mixing length, $\lambda_{\rm conv}/r=(0.5,0.75,1.0)$, with and without nuclear disintegration losses.  The luminosity carried by the post-shock region was calculated using equation (\ref{eq:luminosity_2}) so that $L=L_2$ (the second case in brackets in equation [\ref{eq:luminosity_cases}]) and thus, $v_{\rm sh}(t)$ and $E(t)$ depend only on the ratio $(\lambda_{\rm conv}/r)^4/(\alpha\ell)$.  

From its pre-shock value of $\dot M(t<t_{\rm sh})\approx 0.14\,M_\odot\,\textrm{s}^{-1}$, the accretion rate has dropped to $\dot M=(0.003-0.03)M_\odot\,\textrm{s}^{-1}$ after the first second from shock formation, and to $\dot M=(0.001-0.004)M_\odot\,\textrm{s}^{-1}$ after ten seconds.  The steeper drops occur in the more energetic shocks with larger convective mixing lengths; at such low accretion rates and $\alpha\sim 0.1$, neutrino cooling is negligible compared to viscous heating, at least at $r>r_{\rm ADAF}\gtrsim 0.5\times 10^7\,\textrm{cm}$; this justifies our leaving out of the cooling term in equation (\ref{eq:global_energy_conservation}).  If the gamma ray luminosity of the LGRB prompt emission is controlled by the rate with which material is accreting onto the black hole, then the steep drop in accretion rate associated with shock expansion could explain the termination of the prompt emission \citep{Lindner:10}.

The model with $\lambda_{\rm conv}/r=1.0$ acquires positive energy, and the envelope is unbound and poised to explode at $\sim 5\,\textrm{s}$ after shock formation, when the shock velocity is $\sim 10,000\,\textrm{km s}^{-1}$.  The model with $\lambda_{\rm conv}/r=0.75$ is on track to acquire positive energy after the shock reaches $r_{\rm crit}\sim r_\star$, which are shock radii that our toy model is not designed to handle.  The model with $\lambda_{\rm conv}/r=0.5$ does not appear to evolve toward a globally unbound state, though of course, for a realistic pre-supernova density profile that declines steeply with radius at $r\gtrsim \frac{1}{2} r_\star$, the shock will ultimately emerge from the star and unbind a fraction of its mass.  The latter model is sensitive to nuclear disintegration losses; with $L_{\rm nuc}\approx-8\,\textrm{MeV}\,\dot M/m_p$, the shock seems to stall at $\sim 2\times10^9\,\textrm{cm}$.  

These results indicate that the potential for explosion in collapsar-accretion-powered objects depends critically on the efficiency of convection.  We are not aware of a numerical calibration of the effective convective mixing length $\lambda_{\rm conv}$, if the latter is defined as the mixing length that gives a MLT heat flux equal to the true energy flux carried by convection in the regime, characteristic of supernovae, in which the convective velocities are comparable to the sound speed.  Such a calibration would improve the toy model presented here.  

The 2.5D axisymmetric hydrodynamic simulations of \citet{Lindner:10}, which were carried out with a realistic equation of state \citep{Timmes:00}, developed a fully convective flow in the shock downstream.  These runs were restricted to the domain with cylindrical radii $R>R_{\rm min}=(0.5-2)\times10^3\,\textrm{km}$ and the stress-to-pressure ratio was $\alpha\approx 0.01$. The luminosity carried by the convective envelope was $F_{\rm conv}\sim 0.05\,c_{\rm s} P$, where $c_{\rm s}$ is the sound speed, and the shock velocity is $v_{\rm sh}\sim (0.5-1.5)\times 10^3\,\textrm{km}\,\textrm{s}^{-1}$.  If we artificially set $r_{\rm ADAF}=R_{\rm min}$ and chose $\alpha=0.01$ and $\ell\sim 3\times10^{17}\,\textrm{cm}^2\,\textrm{s}^{-1}$, then our toy model reproduces the relatively low shock velocities in \citet{Lindner:10}. An additional complication not investigated in Lindner et al. is the exothermic and endothermic compositional change in rising and sinking convective cells.

\section{Nucleosynthesis and Nickel}
\label{sec:nucleosynthesis}

The production of an optically bright supernova requires the synthesis of a substantial mass of $^{56}\textrm{Ni}$.  This requires that a substantial mass of the shocked stellar envelope be heated to temperatures $\gtrsim 5\times 10^9\,\textrm{K}$. Also, the reprocessed material must freeze out into iron group elements.  Finally, the proton-to-nucleon ratio during freezeout must be $Y_e\approx 0.5$.  To check whether nickel may indeed be synthesized in accretion-powered explosions, we will examine these requirements, respectively, in Sections \ref{sec:reprocessing}, \ref{sec:freezeout}, and \ref{sec:nickel}, but first, we briefly review some of the different scenarios.  

In the standard model for core collapse supernovae, nickel is synthesized when the shock is fast and the immediate post-shock temperature is sufficiently high \citep[e.g.,][]{Woosley:95,Heger:03}.  Nucleosynthesis calculations for LGRB supernovae and their ultra-energetic version---the hypernovae---typically employ a piston to accelerate supernova ejecta \citep[e.g.,][]{Maeda:09,Dessart:11}, or inject an energetic jet \citep[e.g.,][]{Tominaga:07}, or apply heating in the downstream of the stalled shock \citep[e.g.,][]{Fryer:06} to initiate an explosion. These studies find that high nickel masses inferred in the supernovae associated with LGRBs require the injection of energies $\gtrsim 10^{52}\,\textrm{erg}$ \citep{Tominaga:07,Maeda:09}.  However, the physical mechanism that deposits such large energies in the stellar envelope remains to be elucidated.

In a different scenario, nucleosynthesis in the collapsar scenario occurs in an freely expanding outflow, fireball or wind, coming from a disk of material accreting onto the black hole.  For the outflow to synthesize $^{56}\textrm{Ni}$, the inner disk must be nondegenerate so that proton-neutron equality can be maintained, which is possible with moderate accretion rates; alternatively, neutron-proton equality can be re-established in the wind, and simultaneously, the material must not freeze out in the expanding wind too quickly to produce nickel \citep{Beloborodov:03,Pruet:03,Pruet:04}.  These possibilities are clearly very interesting, but they require the presence of an open funnel through which the outflow from the inner accretion disk can escape.  

A funnel-like density distribution is undoubtedly present at small radii where rotational support is competitive with pressure support.  It is not clear, however, that the funnel can be open at somewhat larger radii, where rotational support is not significant.  The axial region may further be overpressured by the hot cocoon produced during the electromagnetic jet's first traversal of the star.  Lacking a funnel, the disk outflow encounters infalling stellar layers.  In this regime, however, the mechanics of nucleosynthesis in the collapsar must be examined in the context of the interaction and mixing of the outflow with the (shocked) stellar envelope.  To attempt to understand the implications of the interaction of convection-like outflows from the hot inner region with the cooler, but more massive layers of the shocked stellar envelope, we adapt some of the useful approximations developed by \citet{Beloborodov:03} and \citet{Pruet:04}.

\subsection{The Mass Reprocessed to NSE}
\label{sec:reprocessing}

While only a small fraction of the shocked fluid is hot enough to allow nuclear burning, because of the pervasive convection in the shock downstream, a much larger fraction of the shocked fluid can be circulated through the hot inner region and can thus be reprocessed into free nucleons, helium, or the iron group elements.  Let $r_{\rm NSE}$ denote the radius within which NSE among the principal nucleosynthetic products is established on a convective eddy crossing time.  We would like to calculate the mass fraction of the shocked fluid that under the action of the convective mixing visits the radii $r<r_{\rm NSE}$.  For this, we must solve equation (\ref{eq:mixing}) with $X$ denoting the mass fraction of unreprocessed elements subject to the boundary condition that none of the unreprocessed elements survive inside $r_{\rm NSE}$.  We seek a quasi-steady-state solution 
\beq
\label{eq:mixing_steady_state}
\frac{\partial}{\partial r} \left(4\pi r^2 \,\frac{1}{3}\chi_{\rm mix} v_{\rm conv} \lambda_{\rm conv} \rho \frac{\partial X}{\partial r}\right) = 0 .
\eeq
In MLT, the convective velocity can be estimated from equation (\ref{eq:convective_velocity}), which, assuming the power-law pressure and density profiles, $\rho\propto r^{-\delta}$ and $P\propto r^{-\xi}$ and the simple equation of state discussed in Section \ref{sec:eos} reduces to 
\beq
\label{eq:convective_velocity_estimate}
v_{\rm conv}\sim \frac{1}{2}\left(\frac{3}{2}\xi-\frac{1}{2}\delta\right)^{1/2} \,\frac{\lambda_{\rm conv}}{r}\, v_{\rm ff} ,
\eeq
where $v_{\rm ff}$ is the free fall velocity.  At small radii, for the purpose of a rough estimate, we can assume that $v_{\rm ff}\sim (GM_{\rm BH}/r)^{1/2}$ so that $v_{\rm conv}\propto r^{-1/2}$.  If, as before, $\delta=1$ and $\lambda_{\rm conv}\propto r$, we can rewrite equation (\ref{eq:mixing_steady_state}) as 
\beq
\frac{\partial}{\partial r} \left(r^{3/2}\Theta\frac{\partial X}{\partial r}\right) = 0 ,
\eeq
where $\Theta\equiv\frac{4}{3}\pi \chi_{\rm mix} v_{\rm conv}\lambda_{\rm conv} r^{1/2} \rho$ is an approximately radius-independent coefficient.  Integrating this twice and setting $X(r_{\rm NSE})=0$ and $X(r_{\rm sh})=1$, to obtain the mass flux of species $X$ through $r_{\rm NSE}$ is $\dot M_X(r_{\rm NSE}) = \frac{1}{2}\Theta r_{\rm NSE}^{1/2}$.  The fraction $f_{\rm NSE}$ of the shocked fluid that is reprocessed through $r_{\rm NSE}$ then equals
\bea
f_{\rm NSE} &\sim& \frac{\dot M_X(r_{\rm NSE})}{M_{\rm sh}} \frac{r_{\rm sh}}{v_{\rm sh}}\nonumber\\
&\sim& \frac{1}{3} \chi_{\rm mix} \left(\frac{r_{\rm NSE}}{r_{\rm sh}}\right)^{1/2} \frac{\lambda_{\rm conv}(r_{\rm sh})}{r_{\rm sh}} \frac{v_{\rm conv}(r_{\rm sh})}{v_{\rm sh}} .
\eea
\citet{Khokhlov:91} approximated the time scale for convergence to NSE via 
\beq
\label{eq:tau_NSE}
\tau_{\rm NSE}\sim \rho_{\rm g\ cm^{-3}}^{0.2} \exp[179.7/(T_{\rm K}/10^9)-40.5]\,\textrm{s} , 
\eeq
which is consistent with the more recent estimate of \citet{Calder:07}.  Setting $\tau_{\rm NSE}\sim 0.1\,\textrm{s}$, this yields $r_{\rm NSE}\sim (0.5-2)\times10^8\,\textrm{cm}$ where the temperatures are $T_{\rm NSE}\sim 4\times10^9\,\textrm{K}$.  The typical total reprocessed mass prior to the final acceleration of the shock (for $r_{\rm sh}\lesssim r_\star$) for our fiducial toy model is 
\bea
M_{\rm NSE}&\sim& f_{\rm NSE} M_{\rm sh} .
\eea
When at time $t_\star\sim r_\star/v_{\rm sh}(r_\star)$ the shock radius reaches the edge of the stellar model, $r_{\rm sh}\sim r_\star$, the reprocessed mass becomes
\bea
\label{eq:M_NSE_numerical}
M_{\rm NSE}^{t<t_\star}&\sim& 0.15 \,M_\odot\, \chi_{\rm mix} 
\left(\frac{r_{\rm NSE}}{10^8\,\textrm{cm}}\right)^{1/2} \left(\frac{r_\star}{10^{10}\,\textrm{cm}}\right)^{-1} \nonumber\\ &\times&
\left[\frac{\lambda_{\rm conv}(r_\star)}{r_\star}\right]^2\left[\frac{v_{\rm sh}(r_\star)}{5\times10^8\,\textrm{cm}\,\textrm{s}^{-1}}\right]^{-1}
\nonumber\\&\times&  
\left(\frac{M_\star}{10\,M_\odot}\right)^{1/2} \left(\frac{M_\star- M_{\rm BH}}{5\,M_\odot}\right) .
\eea
In deriving equation (\ref{eq:M_NSE_numerical}), we have assumed that the convective velocity is of the form given in equation (\ref{eq:convective_velocity_estimate}).

It is possible that substantial additional nuclear reprocessing inside $r_{\rm NSE}$ can take place over a longer period after the shock has proceeded to accelerate down the steep density gradient of the outer stellar envelope and break out of the star. Thus, the NSE mass estimate quoted in equation (\ref{eq:M_NSE_numerical}) can be considered a lower limit.   If following shock breakout at $\sim t_\star$ the density inside the original stellar radius $r_\star$ decreases exponentially, e.g., on a free fall time $\rho^{t>t_\star}\propto \exp[-v_{\rm ff}(r_\star) t/r_{\rm star}]$, then following shock breakout an additional mass is reprocessed through NSE and can be estimated via
\bea
M_{\rm NSE}^{t>t_\star}&\sim& \int_{t_\star}^\infty \dot M_X(t) dt\nonumber\\&\sim& \frac{r_\star}{v_{\rm ff}(r_\star)} \dot M_X(t_\star)  \exp\left[-\frac{v_{\rm ff}(r_\star)t_\star}{r_\star}\right]\nonumber\\
&\sim&
\frac{v_{\rm sh}(r_\star)}{v_{\rm ff}(r_\star)} \exp\left[-\frac{v_{\rm ff}(r_\star)}{v_{\rm sh}(r_\star)}\right] \,M_{\rm NSE}^{t<t_\star} .
\eea
This shows that if the average shock velocity inside the star is large compared to the free fall velocity at the stellar surface, $v_{\rm sh}\gg\,v_{\rm ff}$, most of the reprocessing to NSE takes place after the shock leaves the star.  To arrive at this conclusion, we have employed a number of extremely crude approximations; a more accurate approach would clearly require carrying out a time integration on a realistic model star.

\subsection{Freezeout into the Iron Group}
\label{sec:freezeout}

What fraction of the reprocessed fluid can turn into iron group elements?  Investigations of the nucleosynthetic footprint of freely expanding winds have been carried out by many authors \citep[e.g.,][]{Beloborodov:03,Pruet:03,Pruet:04,Nagataki:06,Surman:06,Maeda:09,Metzger:11}, but we are not aware of a systematic investigation of nucleosythesis in a  quasi-hydrostatic, convective atmosphere stradling a region in NSE and a frozen-out atmosphere. We anticipate carrying out multidimensional simulations of turbulent convection in the presence of nuclear burning to learn about the compositional yields in such flows.  Here, we attempt to harness the expanding wind solutions by applying them to individual convective cells. This is undoubtely extremely crude, but is consistent in spirit with the nature of the approximations entering the derivation of MLT.

\citet{Pruet:04} calculated the mass fraction of the iron group (more precisely, of $^{56}$Ni since they assume mildly proton-rich conditions, $Y_e=0.51$) in a freely expanding collapsar wind as a function of the entropy per baryon in units of the Boltzmann constant, $S\equiv (m_p/k_{\rm B}) s$ and the variable $\dot M_{\rm wind}/v_{\rm wind}^3$.  Defining the dimensionless parameter $\mu \equiv (\dot M_{\rm wind}/v_{\rm wind}^3) / [0.1\,M_\sun\,\textrm{s}^{-1}/(0.1c)^3]$, an $X_{\rm Fe}>50\%$ freezeout into the iron group requires $\mu\gtrsim(2,13,40)$ for $S=(20,30,40)$, while, similarly, an $X_{\rm Fe}>25\%$ freezeout requires $\mu\gtrsim(0.1,0.6,2,4)$ for $S=(20,30,40,50)$.  The freezeout into iron group prefers low entropies, high densities, and slow convection.

We can apply the \citet{Pruet:04} result to a single rising convective cell by identifying $\dot M_{\rm wind}/v_{\rm wind}^3$ with $4\pi r^2 \rho / v_{\rm conv}^2$.  In the absence of degeneracy, the entropy per baryon at the helium-to-iron group boundary can be estimated from equation (12) in \citet{Pruet:04},
\beq
S\approx5.21\,\frac{T_{\rm MeV}^3}{\rho_{\rm g\, cm^{-3}}/10^{8}}+\frac{1}{4}\left[15.4+\ln\left(\frac{T_{\rm MeV}^{3/2}}{\rho_{\rm g\,cm^{-3}}/10^{8}}\right)\right] ,
\eeq 
which, with $T=T_{\rm NSE}\sim 4\times10^9\,\textrm{K}$, becomes $S\sim21/\rho_6+4.6-0.25\ln\rho_6$, where $\rho_6\equiv\rho(r_{\rm NSE})/10^6\,\textrm{g}\,\textrm{cm}^{-3}$.  Electron degeneracy sets in at temperatures $T_{\rm deg}\lesssim 2\times10^9\,\rho_6\,\textrm{K}$ \citep[e.g.,][]{Beloborodov:03}, below the temperature at which the iron group freezeout occurs. If the velocity of convective cells is as given by equation (\ref{eq:convective_velocity_estimate}) with the exponents given in equation (\ref{eq:power_law_indices}), we obtain 
\beq
\mu\sim  4\, \frac{r_8^3 \rho_6}{M_5} \left[\frac{\lambda_{\rm conv}(r_{\rm NSE})}{r_{\rm NSE}}\right]^{-2} , 
\eeq
where $r_8\equiv r_{\rm NSE}/10^8\,\textrm{cm}$, and $M_5\equiv M_{\rm BH}/5M_\odot$.  This estimate, which may be excessively conservative, suggests that efficient freezeout into iron group elements requires $\rho(r_{\rm NSE})\gtrsim 10^6\,\textrm{g}\,\textrm{cm}^{-3}$.  Such densities are clearly realized in the convective accretion flow, but are probably not realized in the relativistic (or nearly relativistic) axial jet where freezeout is into $\alpha$-particles.

\subsection{Nickel Synthesis and Implications for Supernovae}
\label{sec:nickel}

The proton-to-nucleon ratio of the stellar envelope entering the accretion shock is $Y_e\approx 0.5$, and with this value, $^{56}$Ni dominates the composition of the iron group products produced in the convective accretion flow.  However if significant deleptonization operates at $\sim \textrm{few}\times r_{\rm ISCO}$, convection may transport the neutron-rich fluid near $\sim r_{\rm NSE}$ and thus tip the balance in favor of iron and the lighter iron group isotopes.  \citet{Beloborodov:03} derives an estimate of the equilibrium value of $Y_e(T,\rho)$ in at most mildly degenerate matter that is transparent to neutrinos and applies it to a rotationally-supported accretion flow with vertical scale height $H\sim\frac{1}{2}r$ to conclude that $Y_e$ drops below proton-neutron equality when accretion rates exceed $\dot M> \dot M_n$ where
\beq
\dot M_n = 0.055 \,M_\odot\,\textrm{s}^{-1}\,\left(\frac{\alpha}{0.1}\right)\left(\frac{r}{r_{\rm g}}\right)^{1/2} 
\left(\frac{M_{\rm BH}}{5\,M_\odot}\right)^2 ,
\eeq
where $r_{\rm g}=2GM_{\rm BH}/c^2$.  Our toy model suggests that the accretion rate drops well below $\dot M_n$ very quickly following the initial shock formation, and this implies $Y_e\gtrsim0.5$, where, at densities $\rho\sim 10^8\,\textrm{g}\,\textrm{cm}^{-3}$ characteristic of the innermost disk, $^{56}$Ni dominates the iron group.
%\bea
%Y_e&\sim& \frac{1}{2}\left[1+0.487\,\frac{m_n-m_p}{m_e}\left(\frac{kT}{m_e c^2}\right)^{-1}\right]\nonumber\\& &\times\left[1+1.46 \left(\frac{\hbar}{m_e c}\right)^3 \left(\frac{kT}{m_e c^2}\right)^{-3}\frac{\rho}{m_p}\right]^{-1} .
%\eea

These crude estimates make us optimistic that supernovae powered by collapsar accretion can synthesize nickel masses similar to those required to explain the light curves of supernovae associated with LGRBs.  More detailed work is required to characterize the interplay of convection and nucleosyntesis in the shocked, pressure-supported accretion flow of a collapsar.  A prediction of the present model is that in the supernova ejecta, nickel is mixed with a much larger mass of unburned stellar material. This mixing produces a supernova with a steeper initial rise that is brighter at early times than a spherically-symmetric explosion \citep[see, e.g.,][]{Woosley:06b}.  While our toy model assumes a quasispherical shockwave, the global structure of the explosion should become aspherical just prior to and following shock breakout, and certainly on time scales of $\sim 100\,\textrm{s}$, with higher entropy material outflowing near the rotation axis, as seen in the idealized 2.5D simulations of \citet{Lindner:10}.  Such asphericities can be detected through spectropolarimetry \citep{WangL:08} and spectroscopy \citep[e.g.,][]{Tanaka:09} and if present at shock breakout can also be inferred from the breakout light curve \citep{Couch:11}.  We will address the structure of the ejecta and the implications for the supernova light curve and other observational properties elsewhere.

\section{Conclusions and Discussion}
\label{sec:conclusions}

With the aim of shedding light on the mechanism that produces Type Ic supernovae in LGRB sources, we have developed a toy model for the accretion of a rotating stellar envelope onto a black hole in the aftermath of stellar core collapse.  The purpose of the toy model is to test the ability of collapsar accretion to produce supernovae, and identify aspects of the problem, such as the nature of the ADAF to CDAF transition and the mechanics of convection, that require further investigation. 

The spherically-averaged toy model for a rotating collapsing star assumes that no prompt explosion prior to black hole formation takes place.  The model is constructed to globally conserve mass and energy.  We track the dynamics of the outward traveling shock wave that forms when infalling stellar layers have sufficient angular momentum to be held up the centrifugal barrier and circularize around the black hole. 

The shocked fluid, heated by the dissipation of MHD turbulence that we model with a viscous shear stress term, is convective; we treat this convection in the mixing length approximation.  Some of the dissipated energy is advected into the black hole; the rest is transported by convection through the expanding shocked region and is available to power a supernova.   The amount of energy delivered to the stellar envelope depends on the location of the boundary of the ADAF at small radii and CDAF at large radii.   

The ADAF to CDAF transition is particularly sensitive to the effective convective mixing length. If the mixing length is sufficiently large, our model can acquire positive total energies of at least $\sim 10^{51}\,\textrm{erg}$ over the course of $10$ seconds or longer, which lays open the possibility of a supernova.   It does not seem, however, that the mechanism could produce ``hypernova''-like energies ($\gtrsim 10^{52}\,\textrm{erg}$), at least not in Wolf-Rayet progenitors.

The rate with which shocked stellar fluid accretes onto the black hole drops drastically following the inception of the accretion shock, and thus, losses to neutrino emission are negligible.  If the luminosity of the LGRB prompt emission is correlated with the accretion rate, then the abrupt termination of the prompt emission and the steep decline of the early X-ray afterglow can be interpreted as a consequence of the accretion rate drop \citep[see, also,][]{Lindner:10}.

Because of the rapid convective mixing, tens of percent of a solar mass can be reprocessed through the hot inner radii of the accretion flow where NSE is reached on a dynamical time.  Some reprocessing takes place after the accretion shock breaks out of the star.  Conditions are favorable for the freezing out of the reprocessed matter into $^{56}$Ni, which due to pre-shock-breakout convection should be intermixed with a much larger mass of hydrostatic $\alpha$-elements in the stellar ejecta.

We have assumed throughout that the specific angular momentum of the initial star increases more or less monotonically outward, as one might expect in fully-mixed pre-supernova models \citep{Woosley:06a}.  If this is not the case, then the nonmonotonicity \citep[see, e.g.,][]{Heger:00,Heger:05,Petrovic:05} might have interesting consequences for the evolution of the accretion rate.  For example, the accretion rate may surge if the average specific angular momentum in the shocked region drops below the critical value for rotational support near ISCO, and this might result in a ``flaring'' in the LGRB X-ray light curve \citep[see, e.g.,][]{LopezCamara:10,Perna:10}.

Wolf-Rayet stars that seem to be the most plausible LGRB progenitor candidates were the target this inquiry, but the analysis can be adapted to other contexts in which the collapse of a stellar core into a black hole occurs.  It would be interesting to check whether the collapse into a black hole in the core of a rotating supermassive star \citep[e.g.,][]{Fowler:66,Bond:84,Fuller:86,Baumgarte:99,Shibata:02}, could, as we find here for WR stars, lead to an unbinding of a significant fraction of the remaining stellar envelope.  

In an attempt to elucidate the rapid formation of massive black holes in early galaxies, \citet{Begelman:06,Begelman:08} and \citet{Begelman:10} have proposed that a black hole can form at the center of a large accumulation of gas ($\sim 10^6\,M_\odot$) in a gas-rich primordial galaxy.  The black hole subsequently accretes the gas at the center of the rotating, pressure-supported gaseous object, a ``quasistar,'' in such a way that the object remains gravitationally bound.  In this picture, the accretion rate settles in a quasi-steady state in which the energy dissipated at the innermost radii is transported convectively to the outer radiative zone; the latter, thanks to an internal self-regulating adjustment in the structure of the quasistar, carries a radiative flux that remains below the Eddington limit. Our results suggest that the fate of the quasistar may depend on the energetics of the relatively short period in the immediate aftermath of black hole formation and that the path to a self-regulating quasi-steady state deserves further inquiry.  

\acknowledgements

We would like to thank J. Craig Wheeler and an anonymous referee for comments on the draft.  We acknowledge insightful discussions on the supernova mechanism in GRBs with Lars Bildsten, Adam Burrows, Peter H\"oflich, Davide Lazzati, Andrew MacFadyen, Paolo Mazzali, Ehud Nakar, Tsvi Piran, and Eliot Quataert.  M.~M. acknowledges support from NSF grant AST-0708795 and P.~K. acknowledges support from NSF grant AST-0909110. This material is based upon work supported under a NSF Graduate Research Fellowship awarded to C.~C.~L.

\end{document}